\documentclass[structabstract]{aa}
\usepackage[pdftex,final]{graphicx}
\usepackage{txfonts}
\usepackage{natbib}
\usepackage{multirow}
\usepackage{subfig}
\bibpunct{(}{)}{;}{a}{}{,} 

\begin{document}
\nocite{*}
   \title{Time delays for 11 gravitationally lensed quasars revisited}

\author{E.Eulaers
\inst{1}
\and P. Magain 
\inst{1}
}

\institute{Institut d'Astrophysique et de G\' eophysique, Universit\' e de Li\`ege, All\'ee du 6 Ao\^ut, 17, Sart Tilman (Bat. B5C), Li\`ege 1, Belgium
\email{E.Eulaers@ulg.ac.be}
}

\date{}

  \abstract
   {}
   {We test the robustness of published time delays for 11 lensed quasars by using two techniques to measure time shifts in their light curves.}
   {We chose to use two fundamentally different techniques to determine time delays in gravitationally lensed quasars: a method based on fitting a numerical model and another one derived from the minimum dispersion method introduced by Pelt and collaborators. To analyse our sample in a homogeneous way and avoid bias caused by the choice of the method used, we apply both methods to 11 different lensed systems for which delays have been published: JVAS B0218+357, SBS 0909+523, RX J0911+0551, FBQS J0951+2635, HE 1104-1805, PG 1115+080, JVAS B1422+231, SBS 1520+530, CLASS B1600+434, CLASS B1608+656, and HE 2149-2745}
   {Time delays for three double lenses, JVAS B0218+357, HE 1104-1805, and CLASS B1600+434, as well as the quadruply lensed quasar CLASS B1608+656 are confirmed within the error bars. We correct the delay for SBS 1520+530. For PG 1115+080 and RX J0911+0551, the existence of a second solution on top of the published delay is revealed. The time delays in four systems, SBS 0909+523, FBQS J0951+2635, JVAS B1422+231, and HE 2149-2745 prove to be less reliable than previously claimed.}
   {If we wish to derive an estimate of $H_{0}$ based on time delays in gravitationally lensed quasars, we need to obtain more robust light curves for most of these systems in order to achieve a higher accuracy and robustness on the time delays.}

   \titlerunning{}
   \maketitle

\section{Introduction}

The determination of the time delay between different images of gravitationally lensed quasars is an important step in different kinds of studies: in deriving $H_{0}$, the expansion rate of the Universe \citep[e.g.][]{2008AA...488..481V}, for microlensing studies \citep[e.g.][]{2006AA...455L...1P}, and for detailed studies of the structure of a lensed quasar \citep[e.g.][]{2002MNRAS.334..905G,2008ApJ...676...80M}. 

However, previous time delay determinations have been far from homogeneous, not only because they are based on different methods, but also because of their varying levels of reliability.  Their accuracy depends, among other factors, on the amplitude and shape of the quasar's intrinsic variations, the perturbations of the light curves by microlensing effects, the photometric error bars, the typical time sampling of the monitoring, the total time span of the observing campaign, and the intervals between observing seasons.  Moreover, the published error bars are generally internal errors only and the way in which they are determined varies from study to study. Unfortunately, once a time delay has been published, the value may be used for years without verification, 
even when the authors of the original article caution the reader that the result is not very well-constrained \citep[e.g.][]{2005AA...431..103J}.

Hence, we are of the opinion that it would be useful to re-evaluate published time delays in a number of systems using the same methods as for all of the delays, as well as for the estimate of the error bars.  An idea of the robustness of the time delay values can thus be obtained, not only internally with our methods, but also by comparing our results with published ones. We do not claim that our methods are superior to other ones. However, a critical reanalysis of published results using two fundamentally different approaches allows us to sort the results in terms of the reliability and independence of the method and to determine which lensed systems may be useful for determining $H_{0}$. Our main purpose is thus to examine whether the published light curves allow the determination of reliable time delays. If the answer is positive, we attempt to estimate realistic error bars, and correct some of the published values for small systematic errors.

For the determination of $H_{0}$ by means of gravitational lensing to be competitive with more classical methods \citep[e.g.][]{2009ApJ...699..539R}, we need to reach at least a comparable accuracy of $\sim5\%$ in $H_{0}$. As time delays from different lensed systems should be combined to obtain $H_{0}$, we can assume that the error in the individual time delays contributes to the statistical error in $H_{0}$. Since the time delay uncertainties are only one of several sources of error in $H_{0}$ determinations (to be added e.g. to uncertainties in the dark matter distribution in the lens), they should in any case not exceed $5\%$.

Section 2 and 3 present the two methods used for time delay determination.  These methods are applied to the 11 lensed systems, for which the main results are described in Section 4. The lenses of our sample are those for which accurate astrometry has been determined by means of the deconvolution of near-infrared Hubble Space Telescope images (Sluse et al, submitted to A\&A). A summary of these results is presented in Section 5, together with our conclusions.


\section{Numerical model fit (NMF)}
\label{sec:method}

We revised and improved the method described in \citet{2001AA...380..805B}. The basic idea can be summarized as follows: for a series of given time delays, the method minimizes the difference between the data and a numerical-model light curve with equally spaced sampling points, while adjusting the two parameters of the difference in magnitude between the light curves and a slope that models slow linear microlensing variations. The model is smoothed by introducing the convolution of the model curve with a gaussian $r(t)$ of full width at half maximum comparable to the typical sampling of the observations, and this smoothing term is weighted by a Lagrange multiplier $\lambda$. The function to be minimized is:

\begin{eqnarray}
S &=& \chi^{2} + \lambda \sum_{i}\left(g(t_{i})-(r\ast g)(t_{i})\right)^{2}
\label{eqn:smoothing}
\end{eqnarray}
with 
\begin{eqnarray}
\chi^{2} &=& \sum_{i=1}^{N_{A}}\left(\frac{d_{A}(t_{i})-g(t_{i})}{\sigma_{A}(t_{i})}\right)^{2}\nonumber\\ &+& 
\sum_{i=1}^{N_{B}}\left(\frac{d_{B}(t_{i}-\Delta t)-(\Delta m+\alpha(t_{i}-\Delta t))-g(t_{i})}{\sigma_{B}(t_{i})}\right)^{2},
\label{eqn:chi2}
\end{eqnarray}
as used in \citet{2001AA...380..805B} where $d_{k}(t_{i})$ and $\sigma_{k}(t_{i})$ are the data for image $k  (k=A,B)$ with the associated error bar, $g(t_{i})$ the model curve, $\Delta t$ the time delay, and $\Delta m$ and $\alpha$ the parameters representing the difference in magnitude and slope between the light curves.

The optimal time delay is the one that minimizes the reduced $\chi_{red}^{2}$ between the model and the data points. It is important to insist on the difference between $\chi^{2}$ as defined in Eq. \ref{eqn:chi2} and on the other hand the reduced $\chi_{red}^{2}$

\begin{equation}
\label{eqn:chi2red}
\chi_{red}^{2} = \frac{1}{N_{A}+N_{B}} \chi^{2},
\end{equation}
in which $\chi^{2}$ is divided by the number of data points in common\footnote{By this, we mean the data points lying in the time span for which data for all the lensed images are available after shifting the light curve for the assumed time delay. The number of data points in this common time span is not necessarily the same for the light curve of every lensed image, hence the use of $N_{A}$ and $N_{B}$.} between the light curves of the quasar images for a given time delay. Indeed, the longer the time delay one tests, the fewer points these light curves have in common, which tends to reduce the $\chi^{2}$ and result in a bias towards longer time delays, hence the use of $\chi_{red}^{2}$ to avoid this bias.

A second important difference from the original version of the method is technical: for computational reasons, the length of the model curve should be a power of two, which in some cases proves to be too long in comparison to the data, thus falsifying the balance between data and smoothing terms. In the original version the smoothing term was applied to the full length of the model. We adapted the program in such a way that the part of the model that is only needed to complete the length until the next power of two, is no longer taken into account in the minimization process. In this way, the method becomes independent of the number of data points in the light curve, which was not the case in its original form. 

This method has not only been implemented for two light curves, but also for deriving time delays from three and four light curves \textit{simultaneously}. The strength of this simultaneous approach lies not only in the improved constraints on the model, but also in that we assume the coherence between pairs of time delays, differences in magnitudes, and the slope parameter values.

The robustness of the measured time delay is tested in two ways. First of all, we iteratively attempt to find the three parameters of the model light curve: the spacing of the model curve's sampling points, the range of the smoothing term, and the Lagrange multiplier. The results should be independent of these parameters as long as we remain in a certain range adapted to the data.

In a second step, we wish to test the influence of each individual point of the light curve on the time delay. This is achieved by means of a classical jackknife test: for a light curve consisting of N data points, we recalculate N times the time delay in the light curve of N-1 data by successively leaving out one data point at a time. Time delays should not change drastically because of the removal of a single point from the light curve. If they do, we know which data point is responsible for the change and we can have a closer look at it.

Errors are calculated by means of Monte Carlo simulations. Normally distributed random errors with the appropriate standard deviation are added to the model light curve and the time delay is redetermined. We note that errors are not added to the data as they already contain the observed error, so adding another would bias the results. The model, to which the measurement errors are added, is assumed to provide a more accurate description of the real light curve of the quasar than the data. This procedure is repeated at least 1000 times, preferably on different combinations of smoothing parameters. The mean value of the time delay distribution that we obtain is considered to be the final time delay and its dispersion represents the 1 $\sigma$ error bar. When we have a markedly asymmetrical distribution, we take its mode as the final time delay and use the $68\%$ confidence intervals to obtain error bars. In this paper, all quoted uncertainties are 1 $\sigma$ error bars except where mentioned explicitly.

The advantages of this method are manifold. First, none of the light curves is taken as a reference curve; they are all treated on an equal basis. Second, a model light curve is obtained for the intrinsic variations in the quasar, which is also the case for the polynomial fit method described by \citet{2006ApJ...640...47K}, but not for the minimum dispersion method developed by \citet{1996AA...305...97P}. This is important when calculating the error bars, thus avoiding adding random errors to the data. Finally, since the model is purely numerical, no assumption is made about the quasar's intrinsic light curve, except that it is sufficiently smooth, and we only interpolate the model, never the data.

\section{Minimum dispersion (MD)}

The second method we use is derived from the minimum dispersion method by \citet{1996AA...305...97P} with a number of  adjustments as described in \citet{2010arXiv1009.1473C}. The main improvements to the original \citet{1996AA...305...97P} method consist in:

\begin{enumerate}
 \item No light curve is taken as a reference, they are all treated on an equal basis;
 \item A flexible modelling of microlensing by polynomials up to third order, per light curve or per observing season.
\end{enumerate}

Since no model light curve is constructed, computation time is a lot shorter than for the NMF method. By using two methods based on completely different principles, we are able to check whether the derived time delays are independent of the method, thus testing their robustness (i.e. independence of the particular way in which the data are analysed).


\section{Application to 11 lensed quasars}

We now present the main results of our time delay analysis for each of the published light curves of 11 gravitationally lensed quasars.

\begin{itemize}

 \item \textbf{JVAS B0218+357}
We have used the data set published by \citet{2000ApJ...545..578C}, consisting of 51 flux density measurements at 8.4 GHz and 15 GHz. The authors obtained a time delay $\Delta t_{AB} = 10.1_{-1.6}^{+1.5}$ days, where A is the leading image, thus confirming independently two values published earlier by \citet{1999MNRAS.304..349B} and \citet{1996IAUS..173...37C} of $\Delta t_{AB} = 10.5\pm0.4$ days and $\Delta t_{AB} = 12\pm3$ days, respectively.

\begin{figure}[htbp]
\label{fig:8GHz}\includegraphics[width=\columnwidth]{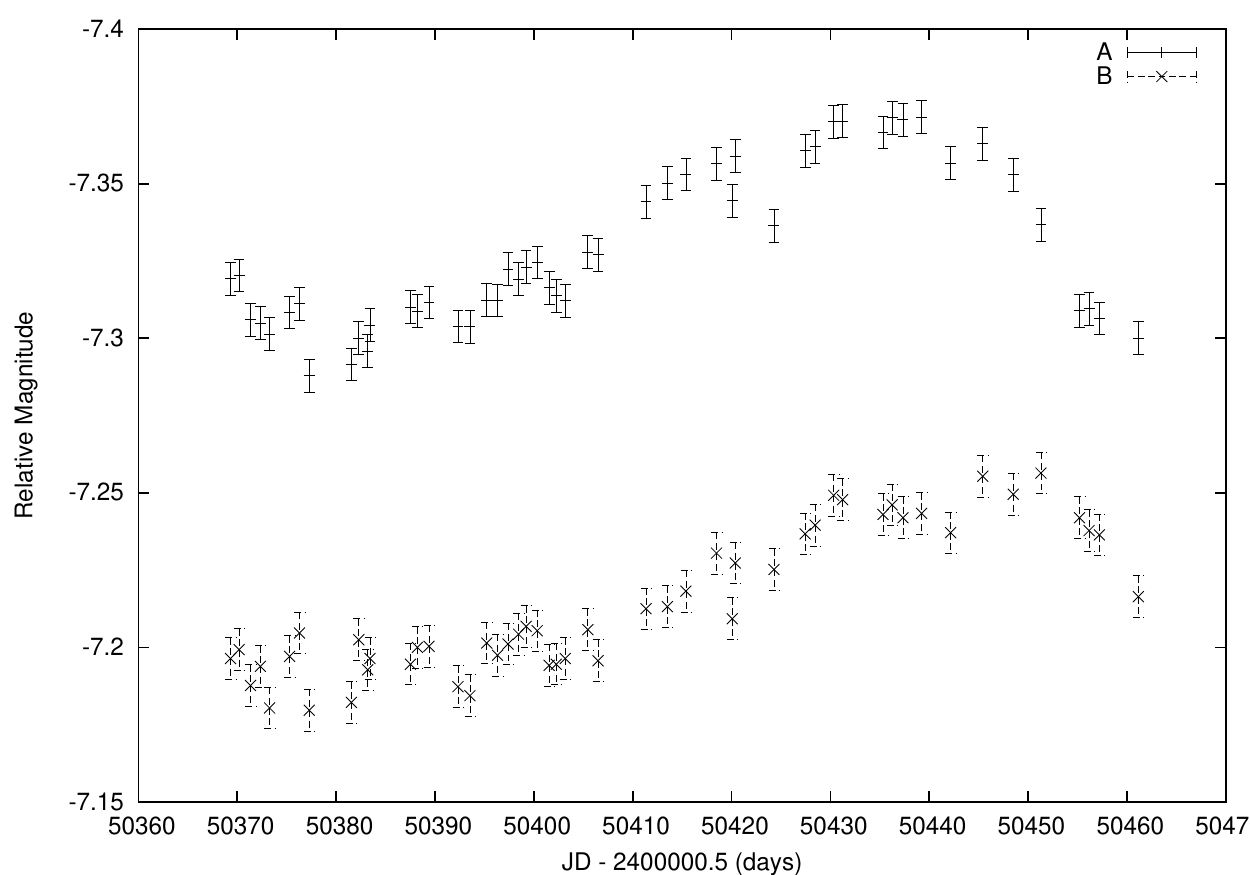}
\label{fig:15GHz}\includegraphics[width=\columnwidth]{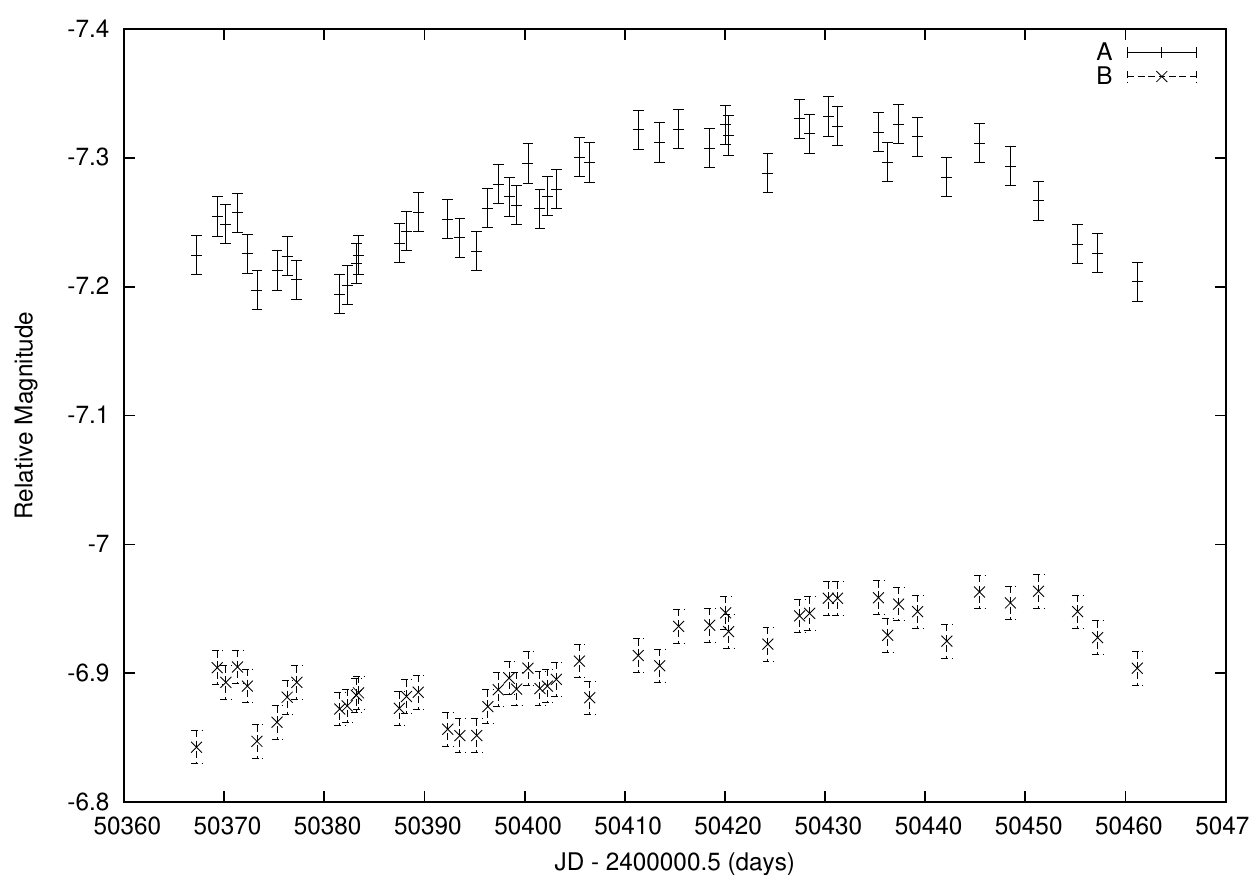}
\caption{Light curves at 8.4 GHz and 15 GHz for JVAS B0218+357 after transforming the flux density measurements into magnitudes.The B curve has been shifted by 1 magnitude for clarity.}
\label{fig:B0218}
\end{figure}

After transforming the flux densities onto a logarithmic scale as shown in Fig. \ref{fig:B0218}, we applied the NMF method to the 8.4 GHz and 15 GHz light curves. Using the entire light curve did not give a clear and unique solution. The jackknife test shows that certain data points can change the value of the time delay. After eliminating three of these points, the 9th, 12th and 35th, from the 8.4GHz light curve, and choosing appropriate smoothing parameters, we obtain a time delay $\Delta t_{AB} = 9.8_{-0.8}^{+4.2}$ days at 68\% confidence level. The larger error bars for higher values of the time delay are due to a secondary peak in the histogram (see Fig. \ref{B0214_histo_8GHz-3pts}) around $\Delta t_{AB} \sim 14$ days. 

\begin{figure}[htbp]
\begin{center}
\includegraphics[width=0.8\linewidth]{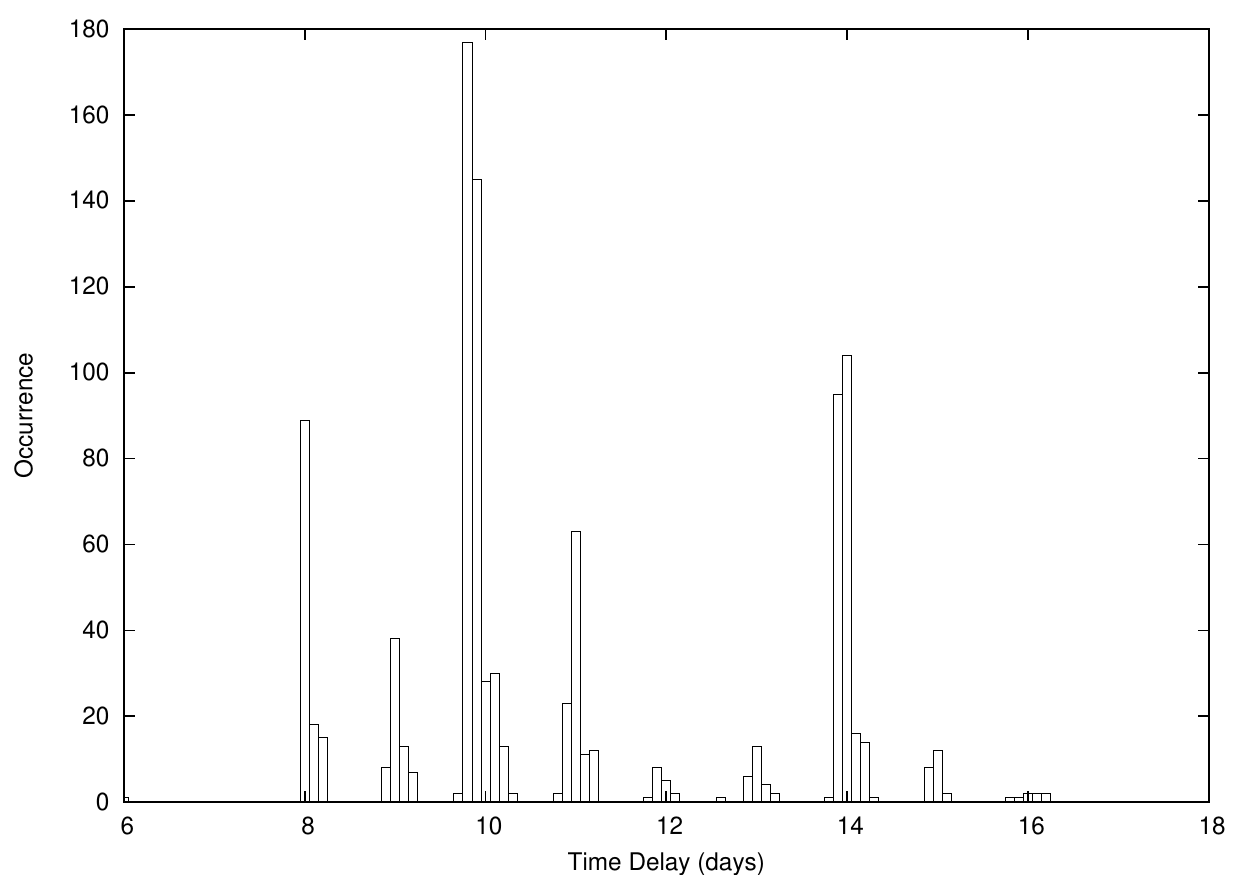}
\caption{Histogram of 1000 runs of the NMF method for the 8.4 GHz data light curve of JVAS B0218+357 leaving out three deviating points. One day gaps in the histogram are artefacts due to the quasi-periodicity of the data.}
\label{B0214_histo_8GHz-3pts}
\end{center}
\end{figure}

Taking into account all points of the 15 GHz light curve provided a comparable value of the time delay of $\Delta t_{AB} = 11.1_{-1.1}^{+4.0}$, even though we noted that the importance of the secondary peak around $\Delta t_{AB} \sim 14$ days was significantly lower after we had eliminated outlying points in both the A and B curve, in the same way as for the 8.4 GHz curve. This suggests that the secondary peak around $\Delta t_{AB} \sim 14$ days is probably caused by artefacts in the data, hence we can confirm with confidence the previously published results: combining the values based on the 8.4 GHz and 15 GHz light curves gives a time delay of $\Delta t_{AB} = 9.9_{-0.9}^{+4.0}$ days.

The MD method confirms the strong influence of these deviating points: the secondary peak around $\Delta t_{AB} \sim 14$ days even completely disappears when they are removed from both the 8.4 GHz and the 15 GHz light curves. The 8.4 GHz curve gives a time delay of $\Delta t_{AB} = 12.6\pm2.9$ days, and the 15 GHz data lead to $\Delta t_{AB} = 11.0\pm3.5$ days, which gives a combined result of $\Delta t_{AB} = 11.8\pm2.3$ days, all in agreement with the above-mentioned values.

Even if all of these time delay values for this object are in agreement with each other, the data do not allow a precision of the order of 5\% in the delays, which would be necessary for a useful estimate of $H_{0}$.

 \item \textbf{SBS 0909+523}
We used the data set published by \citet{2008NewA...13..182G}, which contains 78 data points spread over two observing seasons. Their analysis leads to a time delay $\Delta t_{BA} = 49\pm6$ days where B is the leading image, confirming the previously reported delay $\Delta t_{BA} = 45_{-1}^{+11}$ of \citet{2006AA...452...25U}.

The NMF method, when applied to the entire light curve, gives a delay $\Delta t_{BA} \sim 47$ days, as displayed in Fig. \ref{SBS0909_BA_2-8-400}, which is within the error bars of the previously published delay. On closer inspection however, we note that this delay strongly depends on two points that are outside the general trend of the lightcurve for image B and fall right at the end of the time interval covered by the $A$ data points for this time delay value: the 63rd and the 64th data points. Recalculating the delay while omitting these two points gives a different result of $\Delta t_{BA} \sim 40$ days or even lower values, which is not within the published ranges. 
The same happens if we only take into account the second observing season, which is the longer one: the delay then shortens to $\Delta t_{BA} \sim 40$ days. 
The parameter modelling slow linear microlensing is also significantly smaller in this case.
Visually, both results, with and without the two problematic points, are acceptable.
Nevertheless, although both values have proven to be independent of the two smoothing parameters, the NMF method is sensitive to the addition of normally distributed random errors with the appropriate standard deviation at each point of the model curve. This is because the dispersion in the data points is too small compared to the published error bars. That we obtain $\chi_{red}^{2} \ll1$ also highlights some possible problems in the data reduction or analysis. 

\begin{figure}[htbp]
\begin{center}
\includegraphics[width=0.8\linewidth]{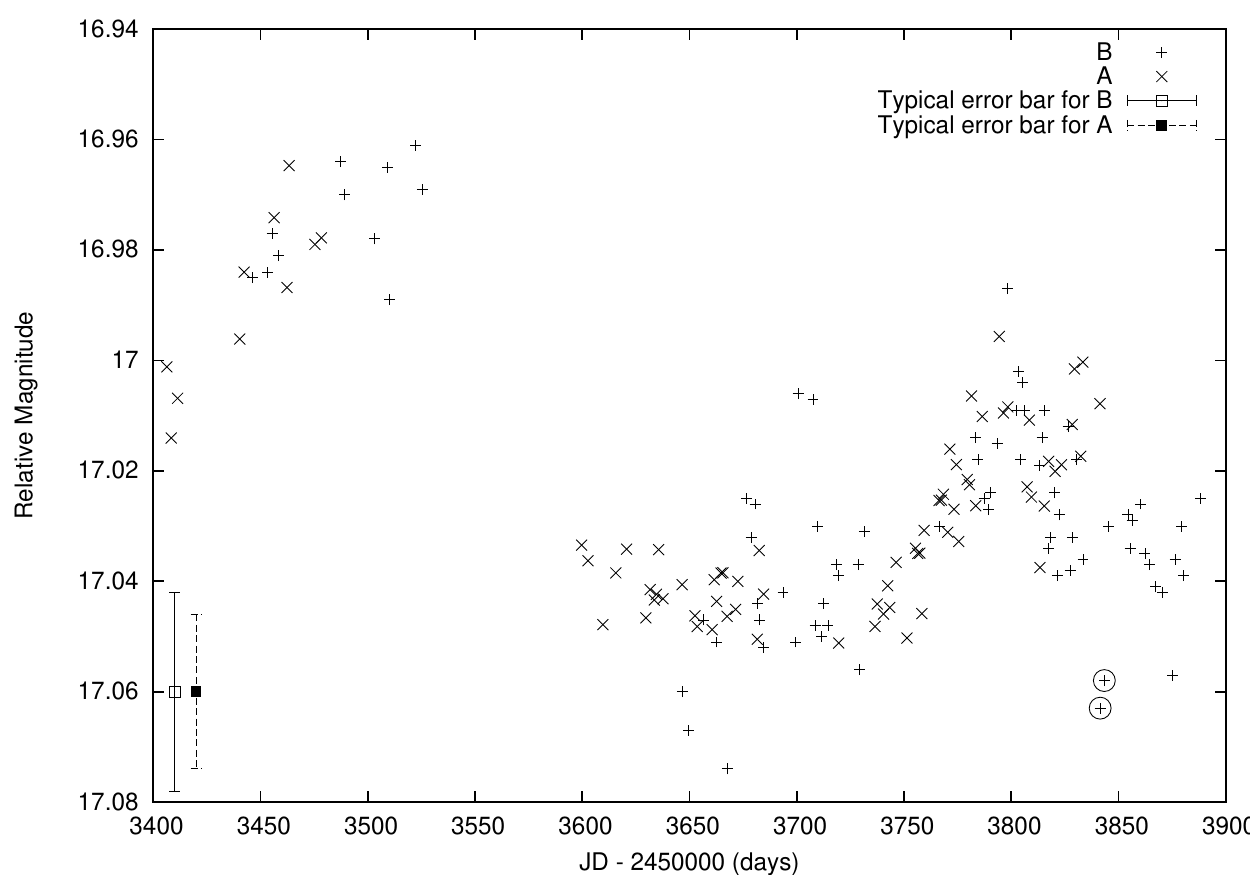} 
\caption{Light curves of SBS0909+523: $A$ is shifted by a time delay $\Delta_t = 47$ days and a difference in magnitude $\Delta m = -0.6656$. The slope parameter $\alpha = 6.0587 \cdot 10^{-5}$ corresponds to a model of slow linear microlensing. The encircled points are the 63rd and 64th observations that were omitted from later tests.}
\label{SBS0909_BA_2-8-400}
\end{center}
\end{figure}

The MD method gives similar results. When using all data points, two possible delays can be seen, depending on the way microlensing is modelled: $\Delta t_{BA} \sim 49$ and $\Delta t_{BA} \sim 36$.
When the two aforementioned data points are left out, we only find $\Delta t_{BA} \sim 36$, independently of how microlensing is handled.

In all cases, with or without these points, leaving more or less freedom for the microlensing parameters, the large photometric error bars result in very large error bars in the time delay when adding normally distributed random errors to the light curves, so that delays ranging from $\Delta t_{BA} \sim 27$ to $\Delta t_{BA} \sim 71$ are not excluded at a 1 $\sigma$ level. 

We conclude that this light curve does not allow a reliable determination of the time delay. 
To determine whether these two data points that do not follow the general trend are due to genuine quasar variations and thus crucial for the time delay determinations or whether in contrast, they are affected by large errors and contaminate the published results, we will need new observations and an independent light curve.
 
\item \textbf{RX J0911+0551}
Data for this quadruply lensed quasar were made available by \citet{2006yCat..34559001P}, but had been previously treated and analysed by \citet{2001PhDT.........4B} and \citet{2002ApJ...572L..11H}, who proposed time delays of $\Delta t_{BA} = 150\pm6$ days and $\Delta t_{BA} = 146\pm4$ days respectively, where B is the leading image of the system and A the sum of the close components A1, A2, and A3. 

\begin{figure}[htbp]
\begin{center}
\includegraphics[width=0.8\linewidth]{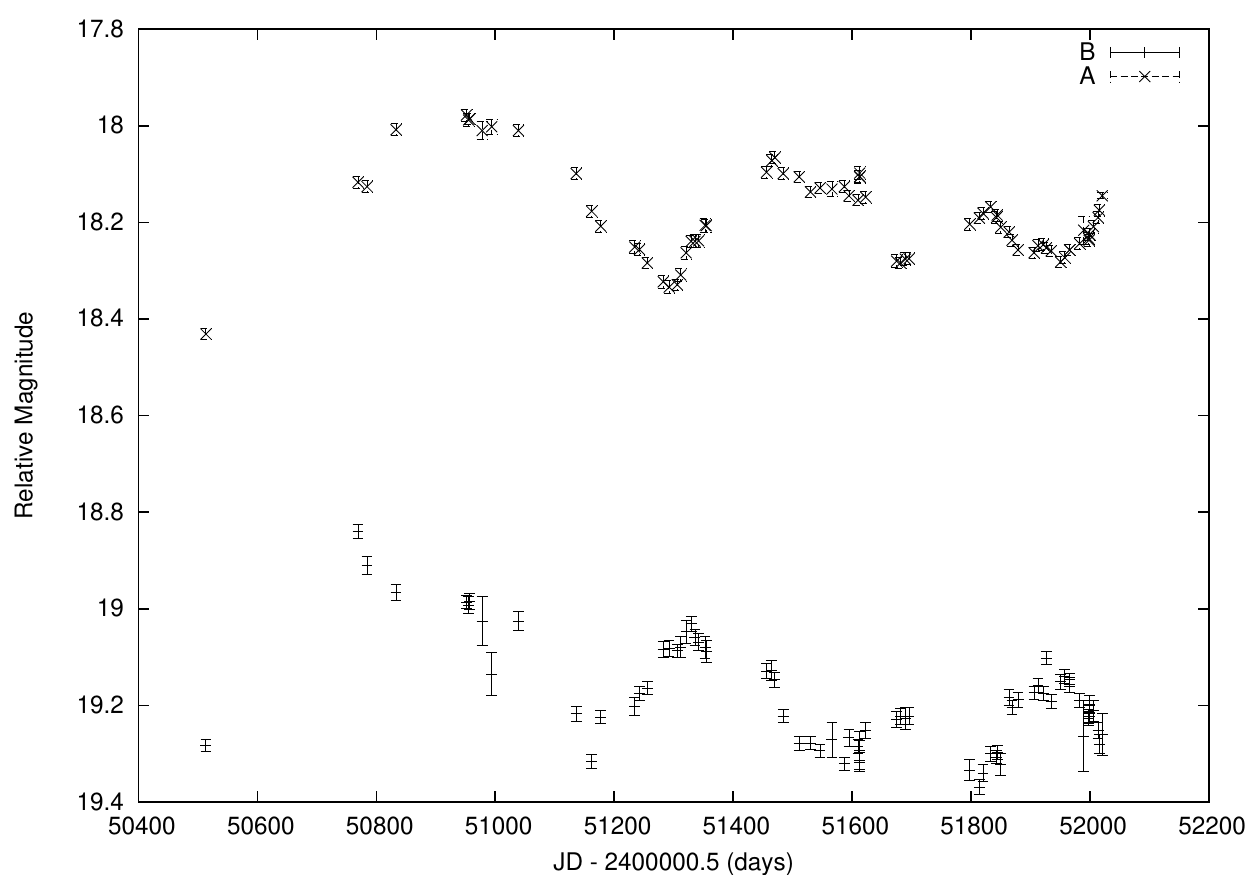} 
\caption{Light curves of RXJ0911+0551: $A$, which is the sum of the close components A1, A2 and A3, has been shifted by one magnitude for clarity.}
\label{fig:RXJ0911}
\end{center}
\end{figure}

Using all data except the first point, which has too strong an influence on our slope parameter because of its isolation as can be seen in Fig. \ref{fig:RXJ0911}, we can at first sight confirm the published delays: the NMF method gives $\Delta t_{BA} = 150\pm2.6$ days and the MD method results in $\Delta t_{BA} = 147.4\pm4.6$ days. However, the histogram in Fig. \ref{RXJ0911+05} shows a secondary peak at $\Delta t_{BA} \sim 157$ days. Investigating this peak further, we come to the conclusion that some points have a very strong influence on the delay: the first observing season, and especially the first ten points of the light curve, indicate a shorter time delay. According to \citet{2001PhDT.........4B}, these points were added to supplement the regular monitoring data of the Nordic Optical Telescope. However, the first three points of the regular NOT monitoring in the B curve are similarly crucial. Omitting these three points leads to larger error bars of $\Delta t_{BA} = 151.6\pm7.0$ days using the NMF method. Finally, recalculating the time delay in the regular monitoring data only and without the first three points in the B curve, gives $\Delta t_{BA} = 159\pm2.4$ days with the NMF method. The MD method results in this case in a histogram with two gaussian peaks, one around $\Delta t_{BA} \sim 146$ days and one around $\Delta t_{BA} \sim 157$ days, implying a mean time delay of $\Delta t_{BA} = 151.4\pm6.7$ days. Only a new and independent light curve of similar length could tell us with more confidence which of these values is correct and which is possibly biased (e.g. by microlensing). 

\begin{figure}[htbp]
\begin{center}
\includegraphics[width=0.8\linewidth]{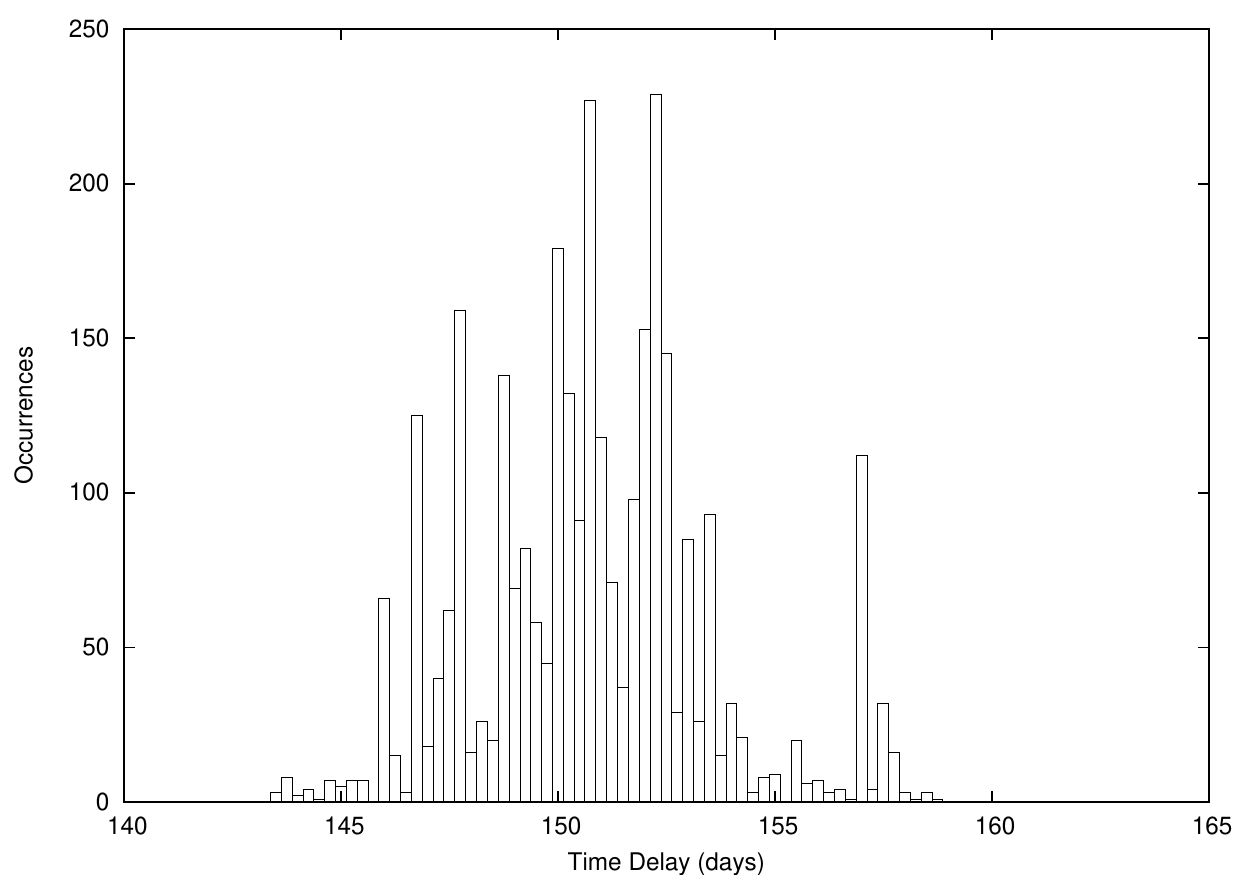}
\caption{Sum of three histograms of 1000 runs each for RXJ0911+05, using three different combinations of smoothing parameters.}
\label{RXJ0911+05}
\end{center}
\end{figure}

 \item \textbf{FBQS J0951+2635}
We used the data set containing 58 points published by \citet{2006yCat..34559001P} and presented in Fig. \ref{fig:FBQ0951}.

\begin{figure}[htbp]
\begin{center}
\includegraphics[width=0.8\linewidth]{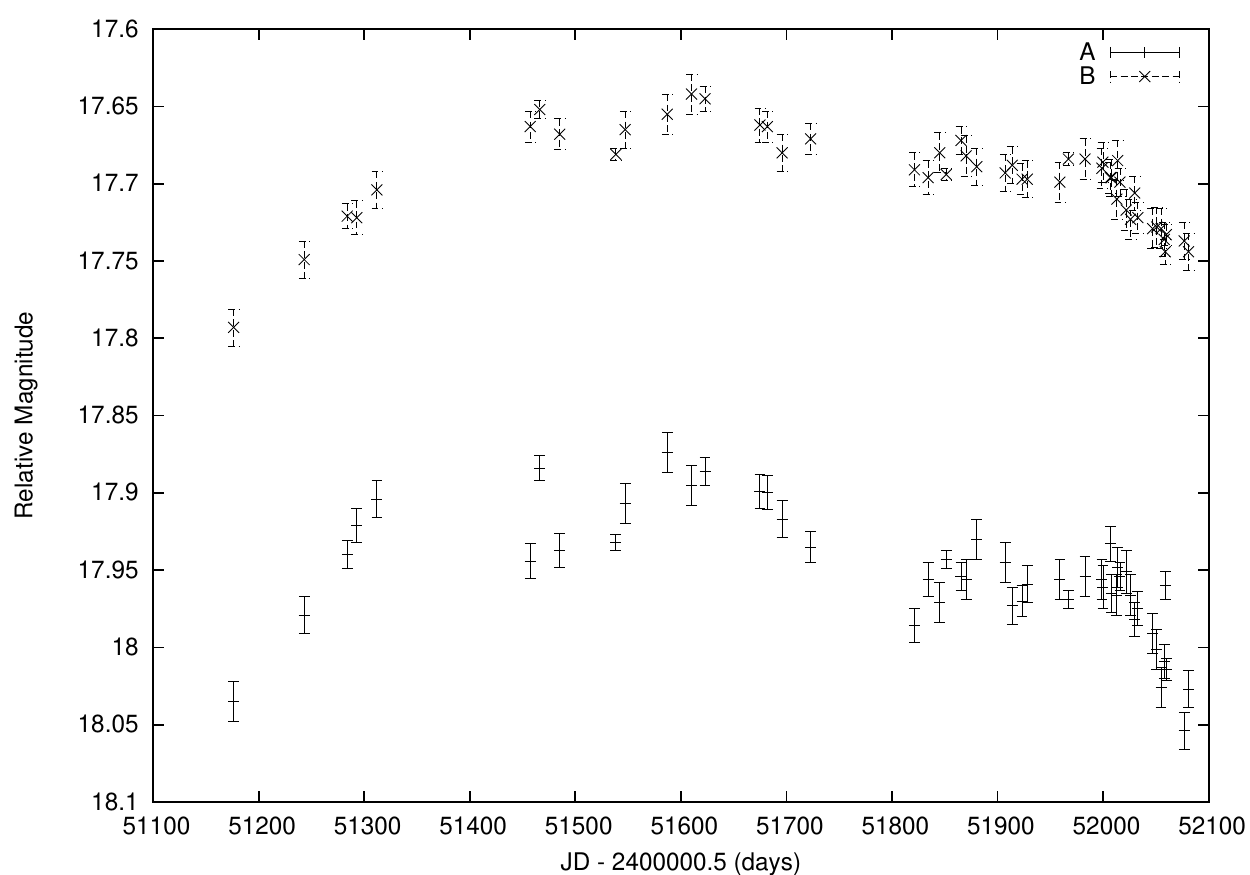} 
\caption{Light curves of FBQ0951+2635. The $B$ curve has been shifted by 0.8 magnitude for clarity.}
\label{fig:FBQ0951}
\end{center}
\end{figure}

\citet{2005AA...431..103J} published a time delay $\Delta t_{AB} = 16\pm2$ days, a result that is only based on the last 38 points of the light curve when the system had been observed more intensively. They found various possible time delays according to the method, the smoothing, and the data points included, so we performed the same tests.

We can confirm that the time delay is very sensitive to the choice of smoothing parameters in the NMF method, especially when using the entire light curve, but is still more sensitive to the data points used: leaving out a single point completely changes the time delay. We calculated time delays in the light curve using between 55 and 58 data points and we found delays ranging from $\Delta t_{AB} = 14.2\pm4.5$ days to $\Delta t_{AB} = 26.3\pm4.7$ days. Taking into account three possible smoothing combinations and four sets of data (leaving out one more data point in each set) leads to a combined histogram of 12000 Monte Carlo simulations, as shown in Fig. \ref{FBQ0951+2635_histo_total}. It is clear that a mean value with error bars $\Delta t_{AB} = 20.1\pm7.2$ days is not of any scientific use: the error bars are too large relative to the time delay. Moreover, the histogram is quite different from a normal distribution. There is no significant concentration of the results, which would allow the determination of a meaningful mode, independently of the chosen binning. One can see that different time delays are possible and can be divided in two groups: shorter values of $\sim10.5$, $\sim15$, and $\sim18.5$ days, and longer values of $\sim26.5$ and $30.5$ days. 

\begin{figure}[htbp]
\begin{center}
\includegraphics[width=0.8\linewidth]{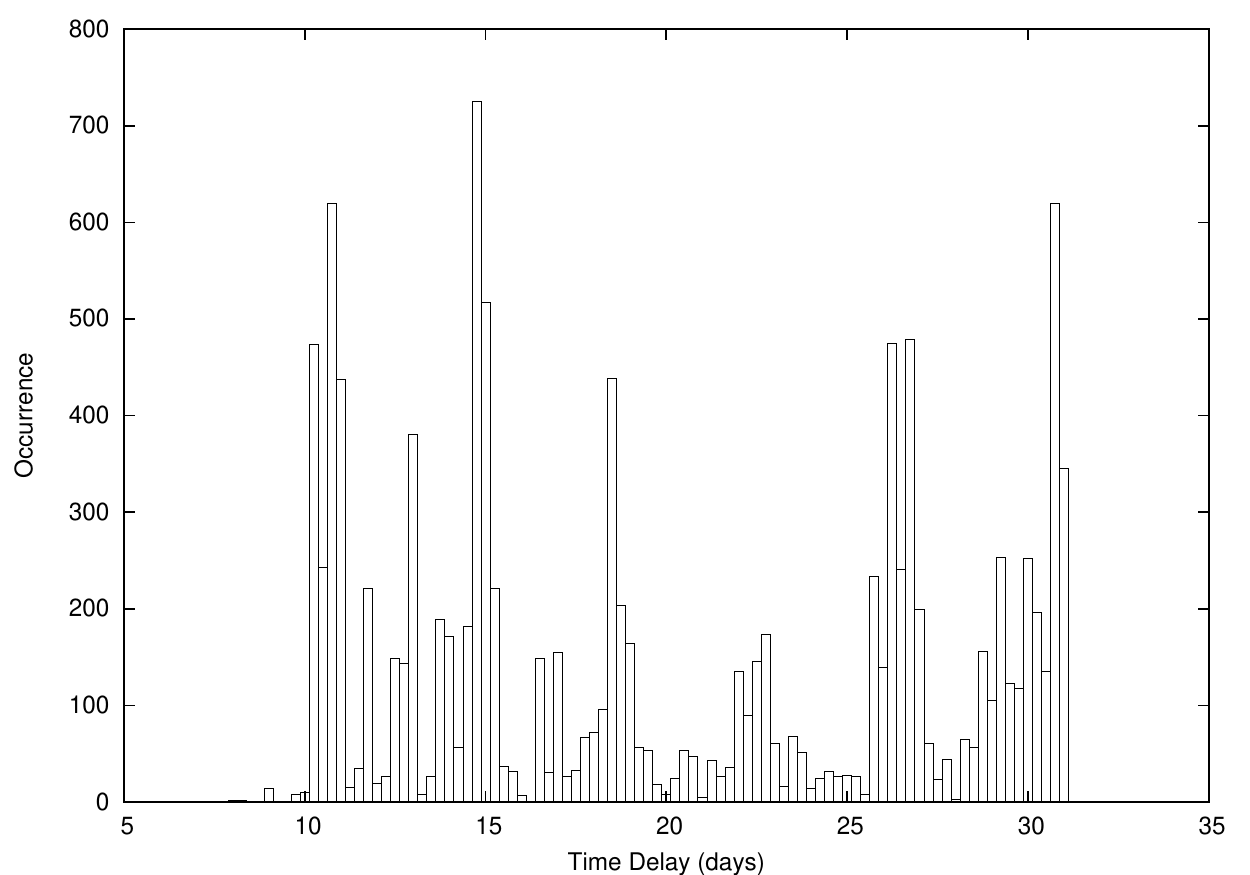}
\caption{Sum of twelve histograms of 1000 runs each for FBQ0951+2635, using three different combinations of smoothing parameters for four sets of data consisting of 58, 57, 56, and 55 data points.}
\label{FBQ0951+2635_histo_total}
\end{center}
\end{figure}

When using only the third observing season, which is more finely sampled, a relatively stable time delay $\Delta t_{AB} = 18.8\pm4.5$ is measured, but once a single point is left out (for example the 19th point of this third season, a point that deviates from the general trend in spite of a small error bar), the result completely changes towards longer values ($\Delta t_{AB} = 25.0\pm4.9$ days) and becomes sensitive to smoothing. As the measured time delay should not depend on the presence or absence of a single point, we can only conclude that this light curve, even if it consists of three observing seasons, does not allow a 
precise determination of this delay. 

The MD method entirely confirms the large uncertainty in this time delay: using all data points we find a time delay of $\Delta t_{AB} = 21.5\pm6.8$ days, whereas the third season only leads to $\Delta t_{AB} = 19.6\pm7.6$ days, with peaks in the histogram around $\Delta t_{AB}\sim12$ and $\Delta t_{AB}\sim28$ days.

According to \citet{1998AJ....115.1371S} and \citet{2005AA...431..103J}, there are spectroscopic indications of possible microlensing, so this might explain the difficulty in constraining the time delay for this system. Longer and more finely sampled light curves might help us to disentangle both effects. However, at the present stage, we can conclude that this system is probably not suitable for a time delay analysis.

 \item \textbf{HE 1104-1805}
We used the data published by \citet{2007ApJ...660..146P}, which combine their own SMARTS R-band data with Wise R-band data from \citet{2003ApJ...594..101O} and OGLE V-band data from \citet{2003AcA....53..229W}. The three data sets are shown in Fig. \ref{fig:HE1104}. Table \ref{tab:TD_HE1104} lists the four time-delay values published for HE1104-1805, where $B$ is the leading image.

\begin{figure}[htbp]
\begin{center}
\includegraphics[width=0.8\linewidth]{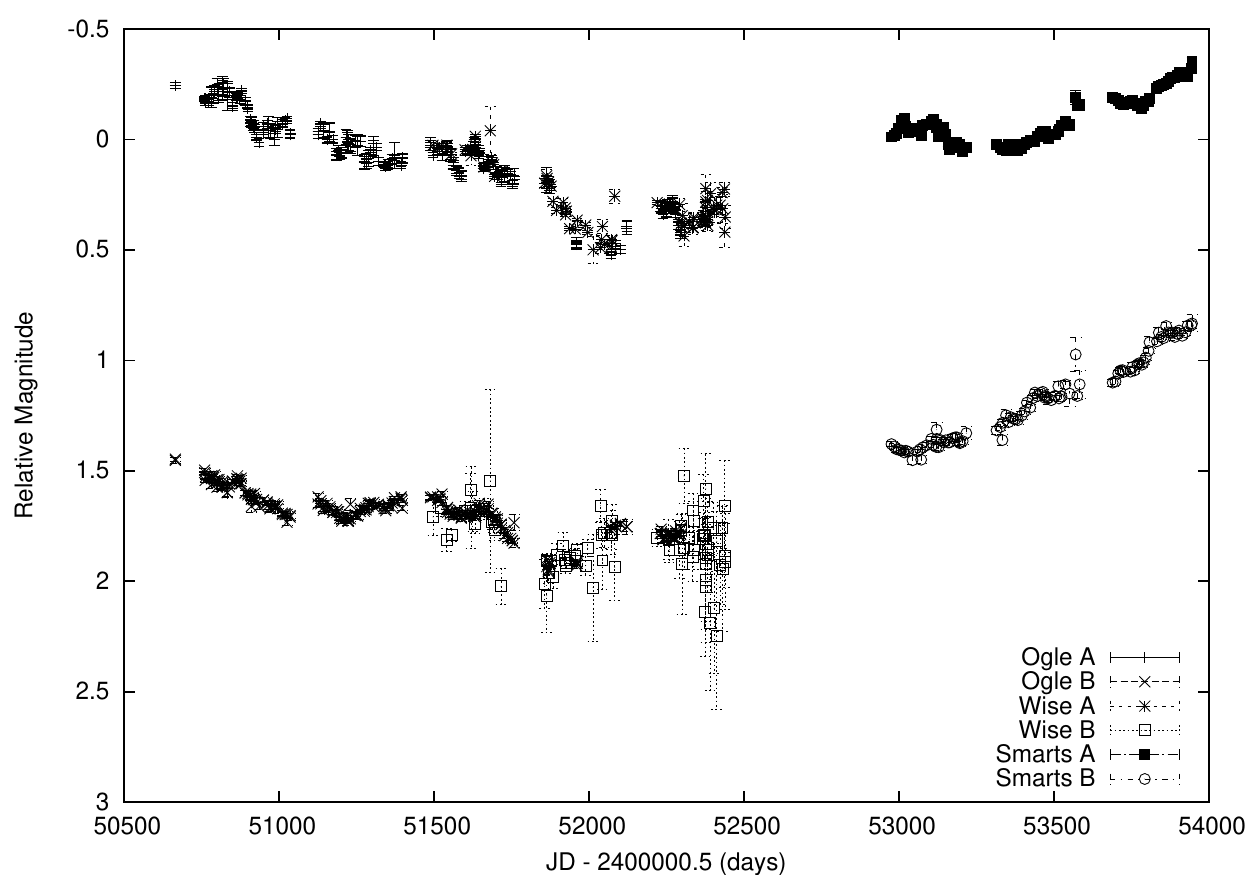} 
\caption{Light curves of HE1104-1805, combining the OGLE V-band data, the Wise R-band data and the SMARTS R-band data.}
\label{fig:HE1104}
\end{center}
\end{figure}

\begin{table}[htbp]
\begin{center}
\begin{tabular}{ll}
Time Delay (days) & Reference \\
\hline
$\Delta t_{BA} = 161\pm7$ & \citet{2003ApJ...594..101O} \\
$\Delta t_{BA} = 157\pm10$ & \citet{2003AcA....53..229W} \\
$\Delta t_{BA} = 152_{-3.0}^{+2.8}$ & \citet{2007ApJ...660..146P} \\
$\Delta t_{BA} = 162.2_{-5.9}^{+6.3}$ & \citet{2008ApJ...676...80M} \\
\end{tabular}
\end{center}
\caption{Published time delays for HE1104-1805.}
\label{tab:TD_HE1104}
\end{table}

We performed tests with both methods on different combinations of the data using all telescopes or only one or two of them. Unfortunately, the results seem to be sensitive to this choice, as they are to the way in which microlensing is treated: both the OGLE and Wise data sets were analysed to find a time delay $\Delta t_{BA} \sim 157$ days, whereas SMARTS data converge to a higher value of $\Delta t_{BA} \sim 161$ days or more as is shown in Fig. \ref{HE1104_OS_AB-2_histo_sum}. In addition, \citet{2007ApJ...660..146P}'s smaller value is recovered with the MD method when including OGLE and Wise data but only for some ways of modelling microlensing. We therefore conclude that we can neither make a decisive choice between the published values, nor improve their error bars, which are large enough to overlap. 

\begin{figure}[htbp]
\begin{center}
\includegraphics[width=0.8\linewidth]{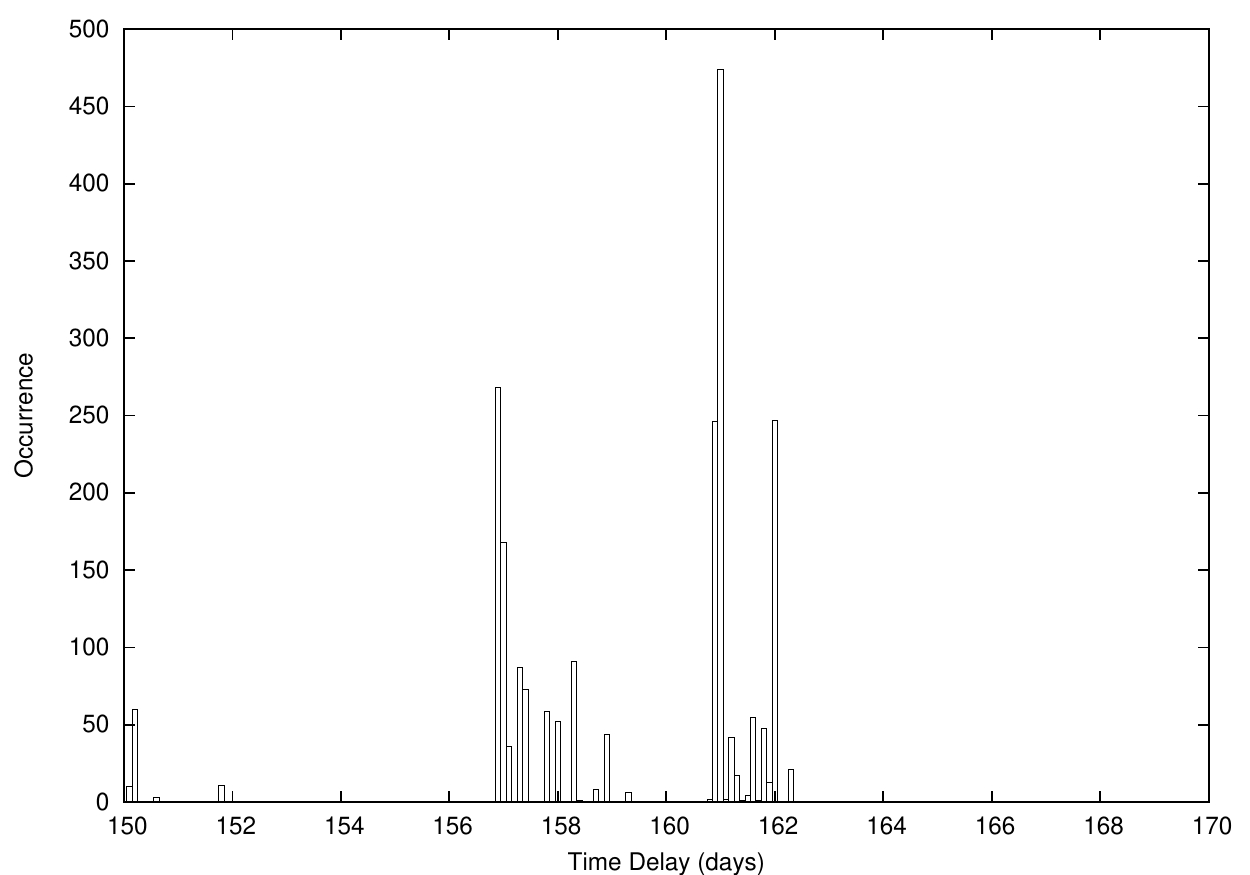}
\caption{Sum of three histograms for HE1104-1805, using three different combinations of smoothing parameters combining only OGLE and SMARTS data. OGLE data point to a time delay $\Delta t_{BA} \sim 157$ days, whereas SMARTS data converge to a longer value of $\Delta t_{BA} \sim 161$ days.}
\label{HE1104_OS_AB-2_histo_sum}
\end{center}
\end{figure}

 \item \textbf{PG 1115+080}
We used the data taken by \citet{1997ApJ...475L..85S}. They published a time delay $\Delta t_{CB} = 23.7\pm3.4$ days between the leading curve $C$ and curve $B$ and estimated the delay between $C$ and the sum of $A1$ and $A2$ at $\Delta t_{CA} \sim 9.4$ days. \citet{1997ApJ...489...21B} used the same data, which are shown in Fig. \ref{fig:PG1115}, but a different method to determine the delays. His value of $\Delta t_{CB} = 25.0_{-3.8}^{+3.3}$ is compatible with Schechter's one. \citet{2008ApJ...689..755M} published new optical light curves for this quadruply imaged quasar in order to study microlensing in the system. Unfortunately, these light curves cannot be used to determine a time delay independently because of the clear lack of features in the variability of the quasar and the inconsistency of the individual error bars relative to the dispersion in the data. 

\begin{figure}[htbp]
\begin{center}
\includegraphics[width=0.8\linewidth]{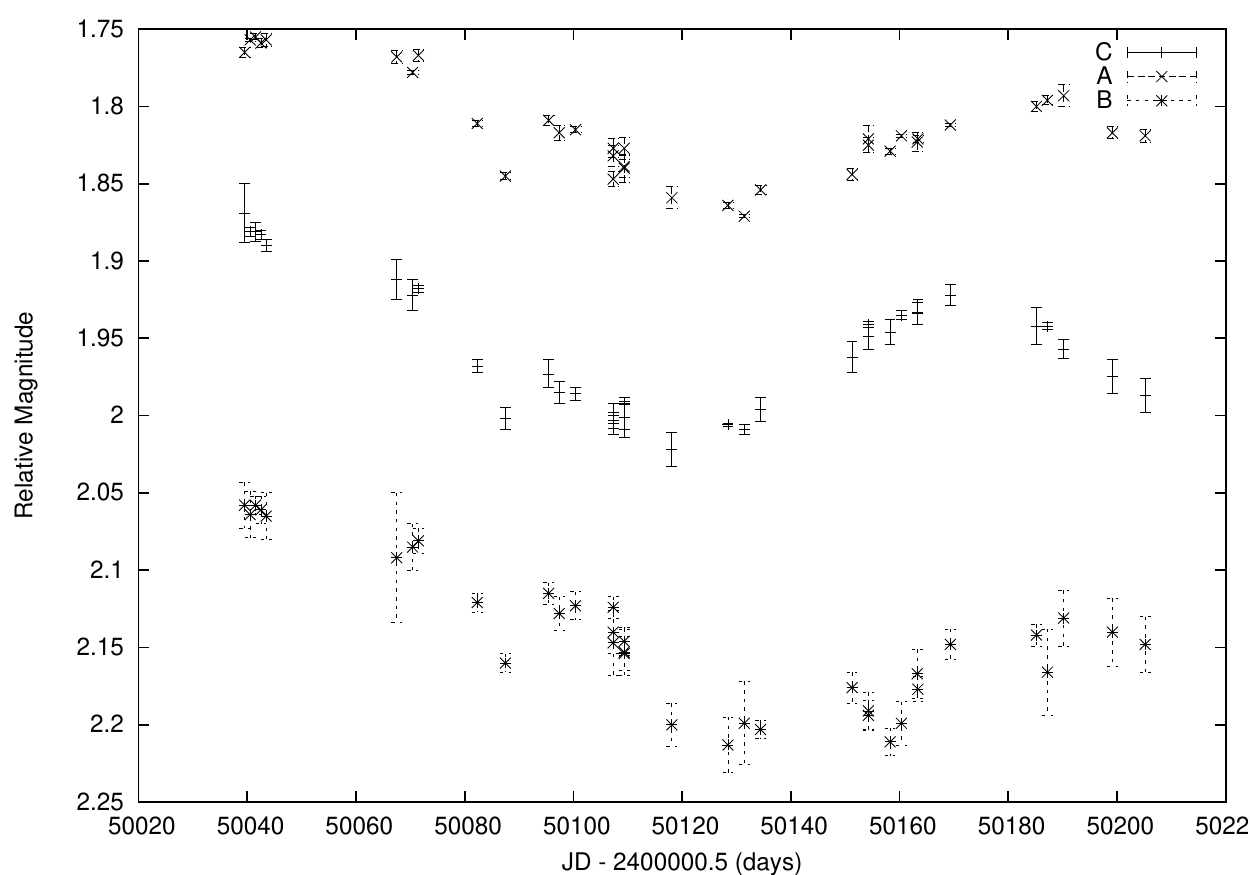} 
\caption{Light curves of PG 1115+080, where the $A$ curve is the sum of the $A1$ and $A2$ component. The $A$ and the $B$ curve have both been shifted by 1.9 and -0.3 magnitude, respectively.}
\label{fig:PG1115}
\end{center}
\end{figure}

A first series of tests on Schechter's data with the NMF method led to time delays of $\Delta t_{CA} \sim 15$ days and $\Delta t_{CB} \sim 20.8$ days with a minor secondary peak around $\Delta t_{CB} \sim 23.8$. We then corrected the published data for the existing photometric correlation between the quasar images and the two stars used as photometric references, as mentioned by \citet{1997ApJ...489...21B}. 
This caused the shorter time delay to shift either towards $\Delta t_{CA} \sim 11$ days or $\Delta t_{CA} \sim 16$ days, and transformed the longer delay into two nearly equally possible results of $\Delta t_{CB} \sim 20.8$ days or $\Delta t_{CB} \sim 23.8$, as indicated by the two main peaks in the histogram in Fig. \ref{PG1115_CB_histosum}. Adding observational errors to the model light curves and taking into account four different ways of smoothing results in $\Delta t_{CA} = 11.7\pm2.2$ (see Fig. \ref{PG1115_CA_histosum}) and $\Delta t_{CB} = 23.8_{-3.0}^{+2.8}$ (see Fig. \ref{PG1115_CB_histosum}).

\begin{figure}[htbp]
\begin{center}
\includegraphics[width=0.8\linewidth]{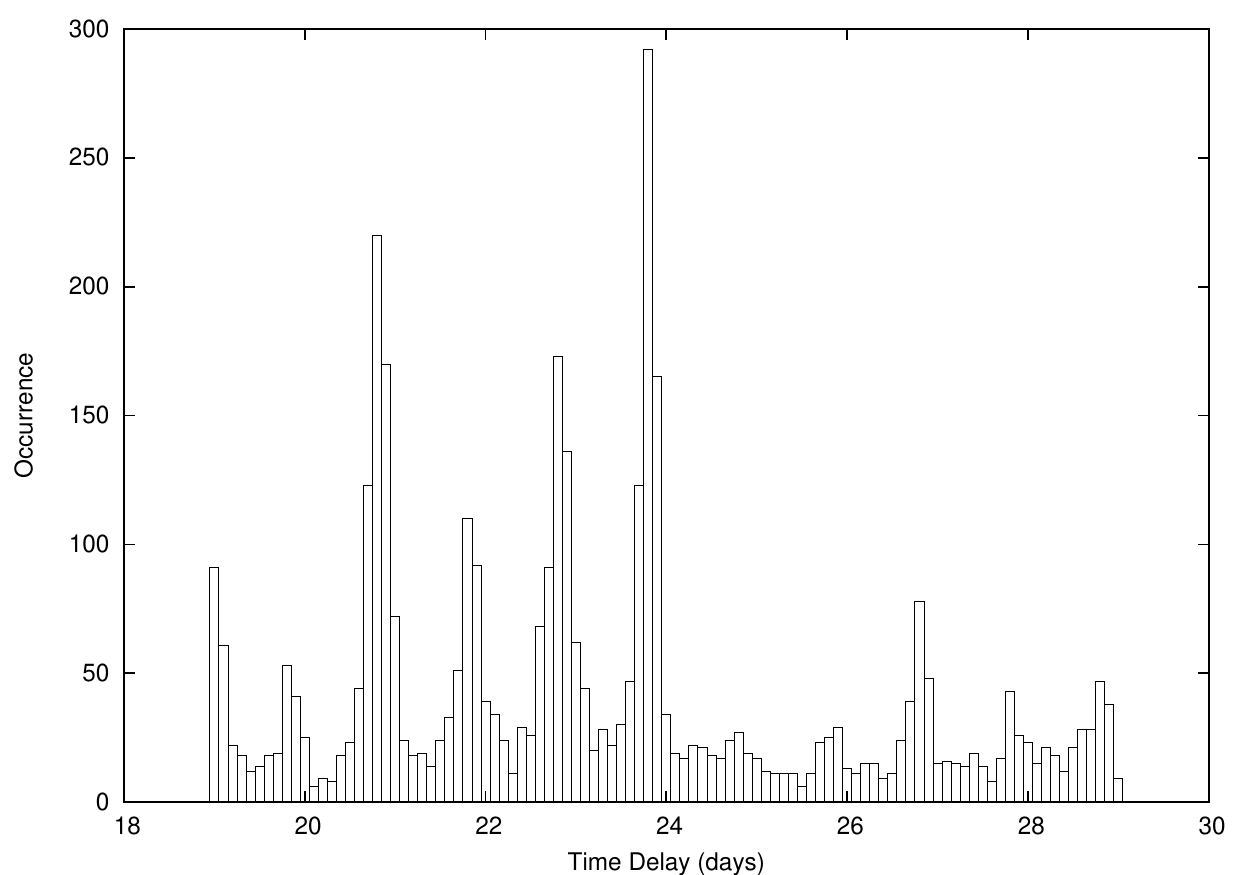}
\caption{Sum of four histograms of 1000 runs each for $\Delta t_{CB}$ in PG1115+080, using four different combinations of smoothing parameters.}
\label{PG1115_CB_histosum}
\end{center}
\end{figure}

\begin{figure}[htbp]
\begin{center}
\includegraphics[width=0.8\linewidth]{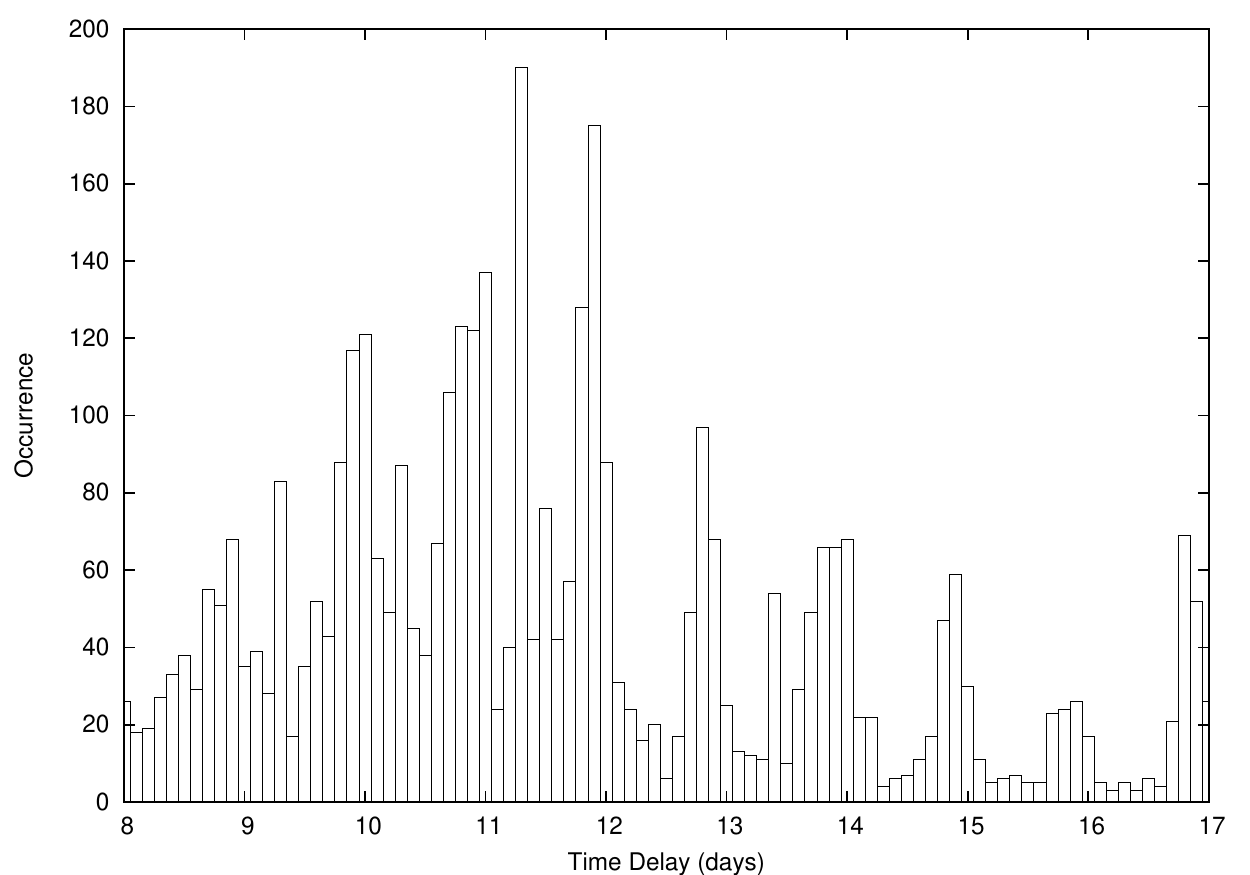}
\caption{Sum of four histograms of 1000 runs each for $\Delta t_{CA}$ in PG1115+080, using four different combinations of smoothing parameters.}
\label{PG1115_CA_histosum}
\end{center}
\end{figure}

The MD method confirms a delay of $\Delta t_{CB} \sim 20.0$ days, but finds a second solution around $\Delta t_{CB} \sim 12.0$ days, which results in $\Delta t_{CB} = 17.9\pm6.9$. The value for $\Delta t_{CA}$ is also shorter than the one obtained with the other methods namely $\Delta t_{CA} = 7.6\pm3.9$ days and has larger error bars. Unfortunately, the length and quality of this light curve do not allow one to choose between the possible time delays that differ according to the method used, but are generally lower than published values.

 \item \textbf{JVAS B1422+231}
For this quadruply lensed quasar, we used the data published by \citet{2001MNRAS.326.1403P}, consisting of flux density measurements at two frequencies, $8.4$ and $15$ GHz. Their results for the time delays were based only on the $15$ GHz data without image $D$, which is too faint. These data are shown in Fig. \ref{fig:B1422}. They obtained $\Delta t_{BA} = 1.5\pm1.4$ days, $\Delta t_{AC} = 7.6\pm2.5$, and $\Delta t_{BC} = 8.2\pm2.0$ days when comparing the curves in pairs. 

\begin{figure}[htbp]
\begin{center}
\includegraphics[width=0.8\linewidth]{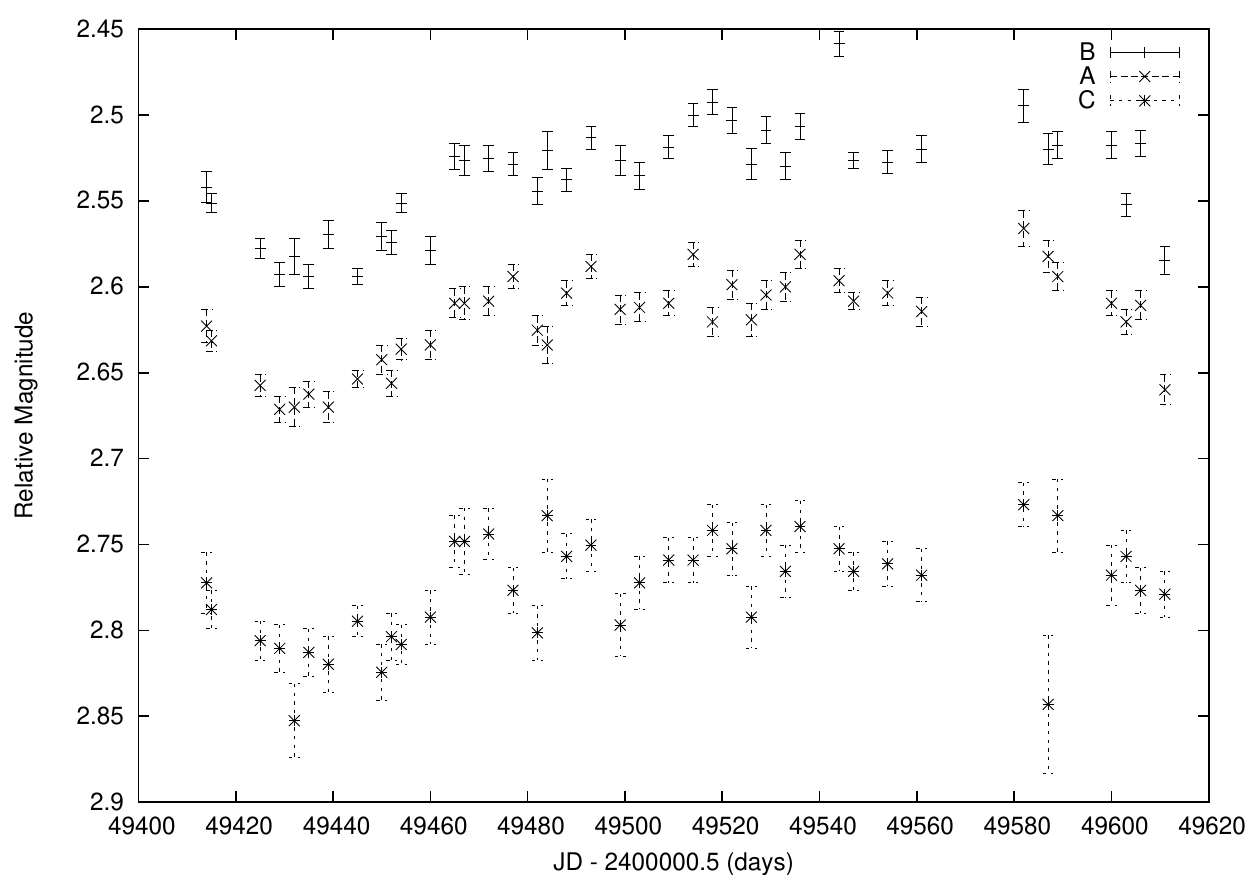} 
\caption{Light curves of JVAS B1422+231 for the $A$, $B$, and $C$ components after transforming the flux density measurements at the $15$ GHz frequency into magnitudes. The $C$ curve has been shifted by half a magnitude.}
\label{fig:B1422}
\end{center}
\end{figure}

The NMF method allows time delays to be tested for the three light curves simultaneously, thus imposes coherence on the results. Given the error bars in the published results, which are large compared to the time delays, we can confirm the results here, but we emphasize that they include two distinct groups of solutions between which we cannot decide based on the actual light curves: for the shortest delay, we either have $\Delta t_{BA} \sim 1.0$ day or $\Delta t_{BA} \sim 2.0$ days, as shown in Fig. \ref{B1422_BA_tot_sum}. The choice between both solutions is sensitive to the smoothing parameters: the importance of the first group $\Delta t_{BA} \sim 1.0$ day is lower, and even disappears completely, with greater smoothing. 

\begin{figure}[htbp]
\begin{center}
\includegraphics[width=0.8\linewidth]{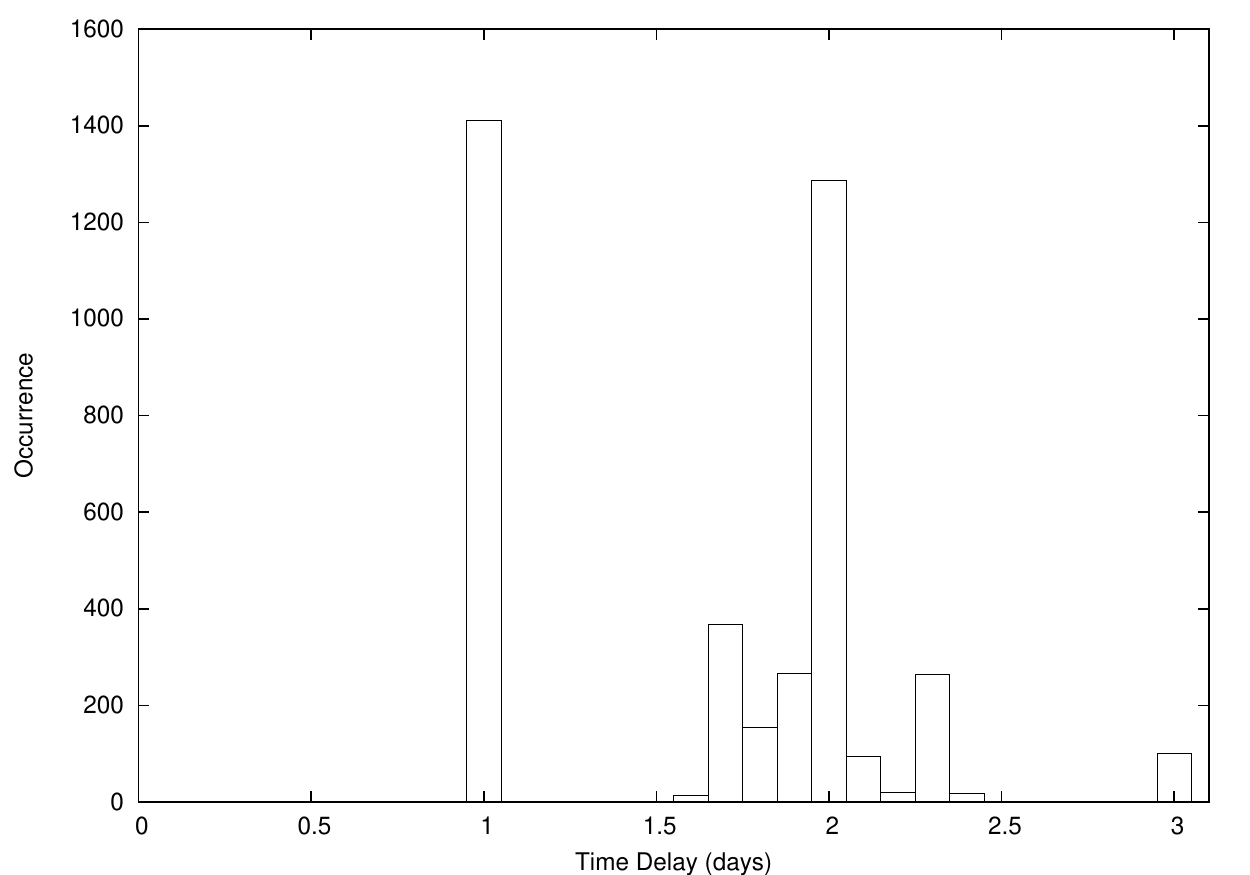}
\caption{Sum of four histograms for $\Delta t_{BA}$ of 1000 runs each for JVAS B1422+231, using four different combinations of smoothing parameters.}
\label{B1422_BA_tot_sum}
\end{center}
\end{figure}

For $\Delta t_{BC}$, the situation is similar but even less clear: low smoothing parameters lead to a range of possible solutions between $\Delta t_{BC} \sim 6$ and $\Delta t_{BC} \sim 10$ days, which are all within the error bars of the published results. However, Monte Carlo simulations of reconstructed light curves with higher smoothing parameters give a time delay $\Delta t_{BC} = 10.8\pm1.5$ days with a secondary peak around $\Delta t_{BC} \sim 8$ days. 

The MD method gives a completely different result: it converges towards time delays that invert the BAC-order into CAB but with error bars large enough not to exclude the BAC order of $\Delta t_{BA} = -1.6\pm2.1$ days, $\Delta t_{AC} = -0.8\pm2.9$, and $\Delta t_{BC} = -2.4\pm2.7$ days.

New observations are clearly necessary to reduce the uncertainties in both the different solutions and the error bars that are too large in comparison with the time delays to be useful to any further analysis.

 \item \textbf{SBS 1520+530}
Two data sets exist for this doubly lensed quasar: the set made available by \citet{2002yCat..33910481B} and the one published by \citet{2005yCat..34400053G}. \citet{2002AA...391..481B} were the first to publish a time delay for this system $\Delta t_{AB} = 128\pm3$ days, where $A$ is the leading image, or $\Delta t_{AB} = 130\pm3$ days when using the iterative version of the method \citep{2001AA...380..805B}. \citet{2005AA...440...53G} used an independent data set and found four possible time delays, of which the one with the largest statistical weight, $\Delta t_{AB} = 130.5\pm2.9$, is perfectly consistent with the previously published time delays.

Even if the light curve based on \citet{2005yCat..34400053G} data contains more than twice as many data points as \citet{2002yCat..33910481B}'s older light curve, we decided not to use it because of the lack of overlapping data between the $A$ and the $B$ curves of the quasar after shifting the $B$ curve for the time delay, as can be seen in Fig. \ref{SBS1520-Gayn}.

\begin{figure}[htbp]
\begin{center}
\includegraphics[width=0.8\linewidth]{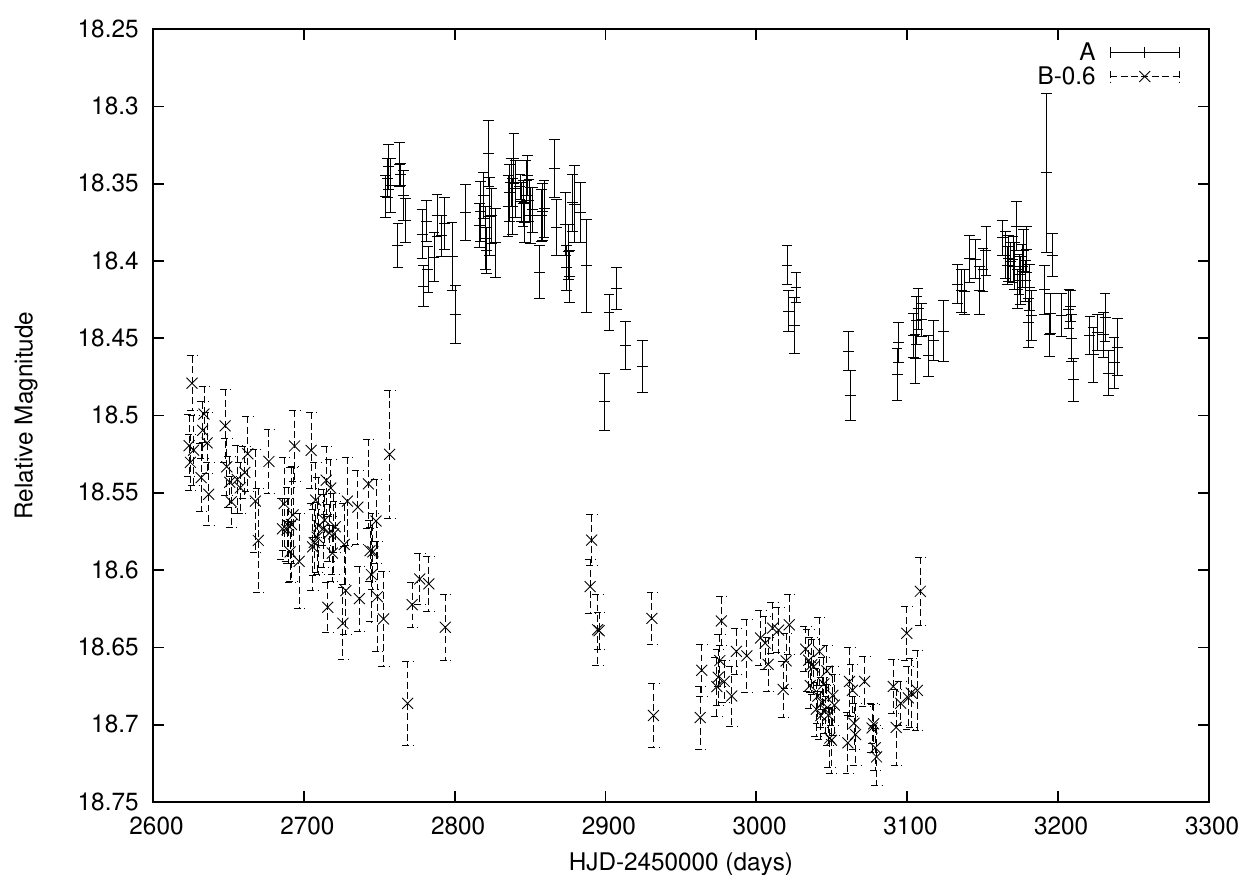}
\caption{Light curves for SBS 1520+530 based on \citet{2005yCat..34400053G}'s data after shifting the $B$ curve by $\Delta t_{AB} = 130.5$ days. Hardly any data points overlap between the light curves for images $A$ and $B$.}
\label{SBS1520-Gayn}
\end{center}
\end{figure}

Applying different tests to \citet{2002AA...391..481B}'s light curves, shown in Fig. \ref{fig:SBS1520}, using the NMF method led to a time delay for which the error bars overlap with the published value: $\Delta t_{AB} = 126.9\pm2.3$. That the delay is slightly shorter than \citet{2002AA...391..481B}'s value can be explained by our use of the reduced $\chi_{red}^{2}$ instead of the $\chi^{2}$, the latter implying that longer delays are the more likely ones, as explained in Section \ref{sec:method}. This effect was also noted using the iterative method. 

\begin{figure}[htbp]
\begin{center}
\includegraphics[width=0.8\linewidth]{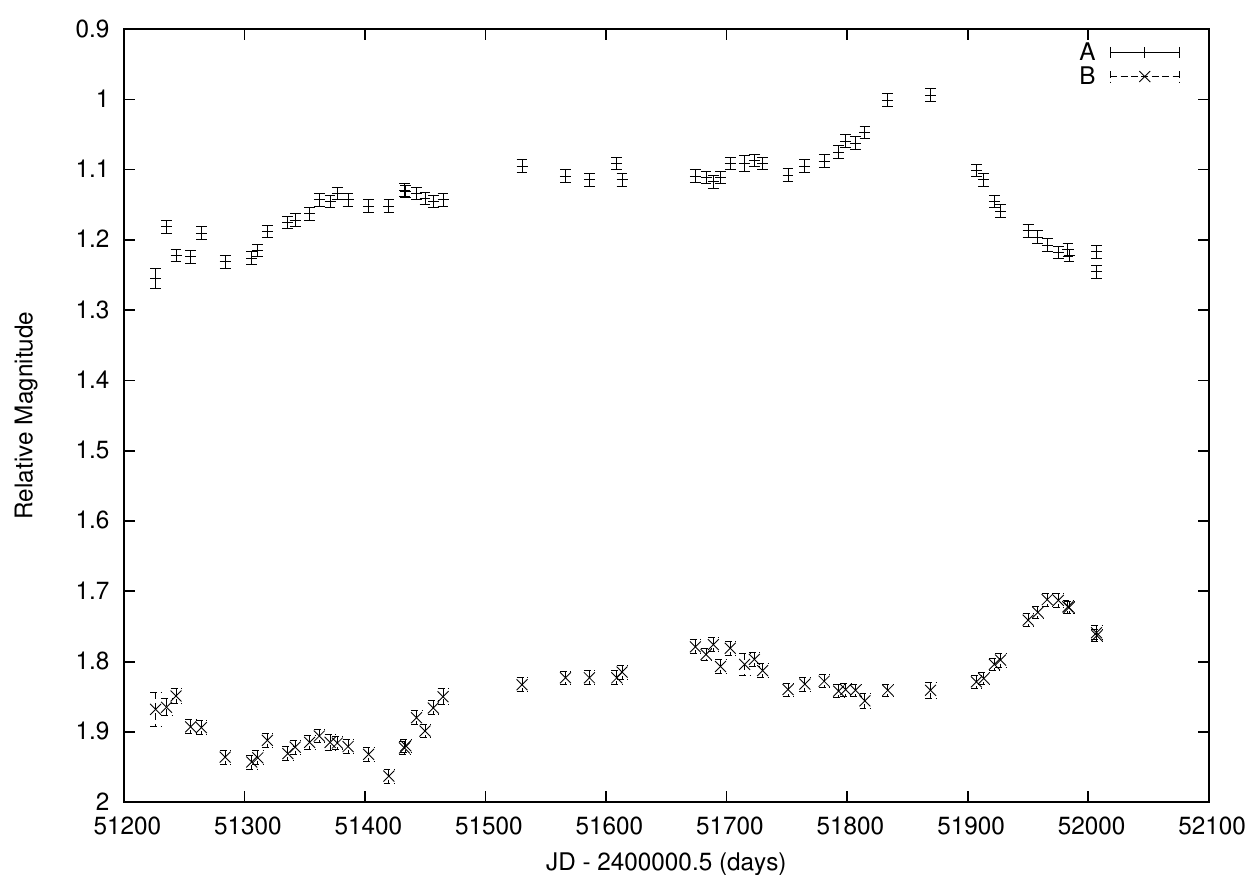}
\caption{Light curves for SBS 1520+530 based on \citet{2002AA...391..481B}'s data.}
\label{fig:SBS1520}
\end{center}
\end{figure}

\begin{figure}[htbp]
\begin{center}
\includegraphics[width=0.8\linewidth]{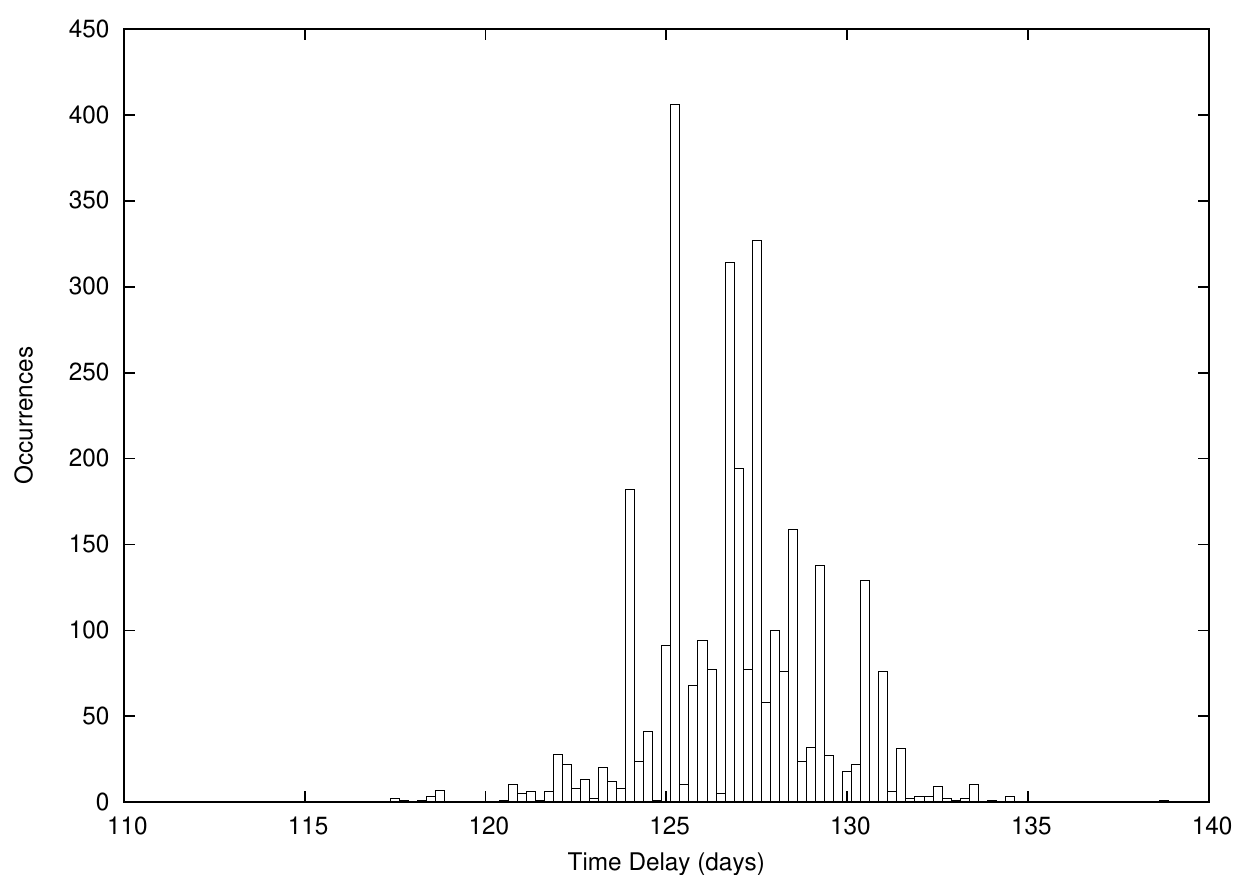}
\caption{Sum of three histograms of 1000 runs each for SBS1520+530, using three different combinations of smoothing parameters.}
\label{SBS1520_histo_somme}
\end{center}
\end{figure}

The MD method yields a comparable time delay of $\Delta t_{AB} = 124.6\pm3.6$ days and confirms the shape of the histogram: the highest peak value at $\Delta t_{AB} \sim 125$ days and a clear secondary peak at $\Delta t_{AB} \sim 127.5$ days. Combining the values from both methods implies that $\Delta t_{AB} = 125.8\pm2.1$ days.

 \item \textbf{CLASS B1600+434}
Data, as shown in Fig. \ref{fig:B1600}, were made available by \citet{2006yCat..34559001P} but had been treated and analysed by \citet{2000ApJ...544..117B}, who published a final time delay $\Delta t_{AB} = 51\pm4$ days, where $A$ is the leading image, consistent with the time delay $\Delta t_{AB} = 47_{-9}^{+12}$ days from \citet{2000AA...356..391K} based on radio data.

\begin{figure}[htbp]
\begin{center}
\includegraphics[width=0.8\linewidth]{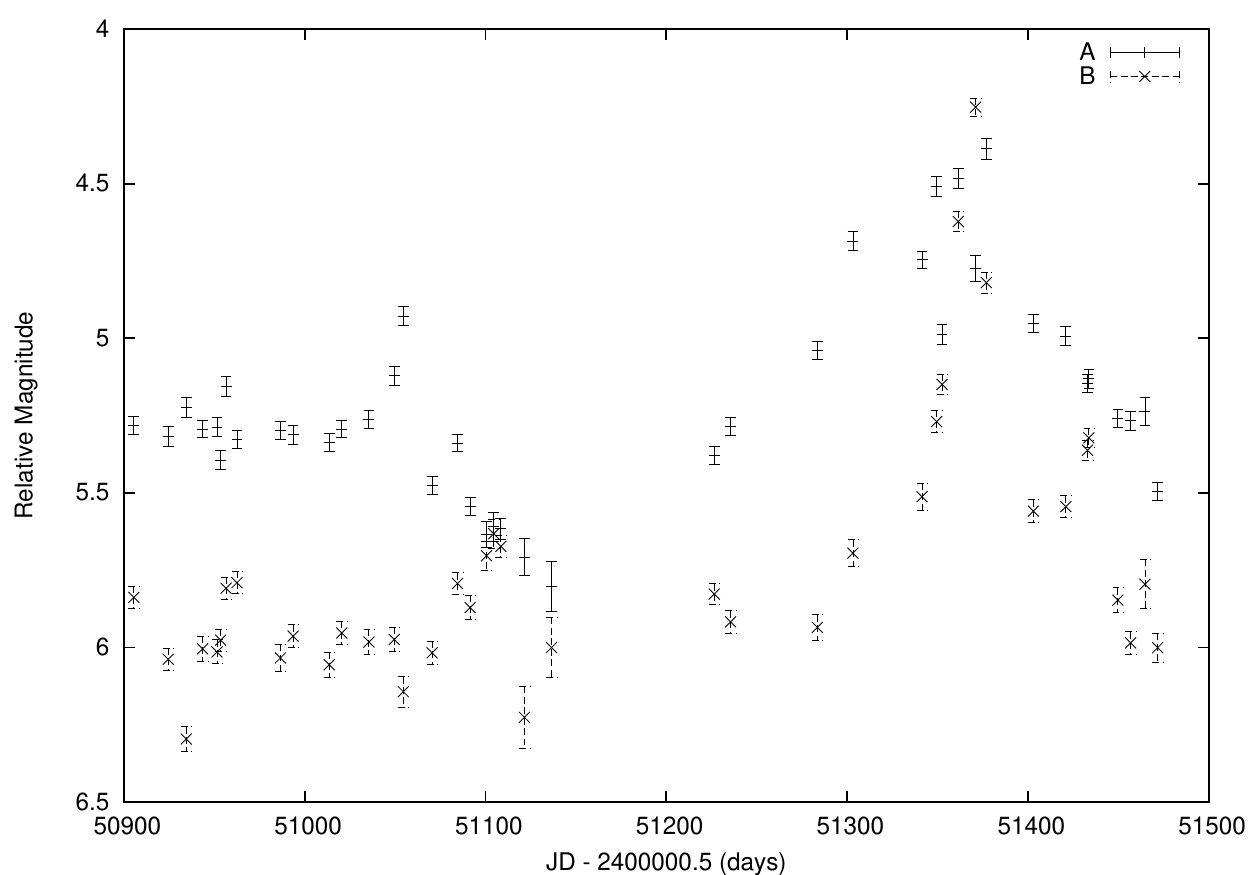}
\caption{Light curves for CLASS B1600+434: $41$ data points spread over nearly two years.}
\label{fig:B1600}
\end{center}
\end{figure}

The NMF method leads to a time delay $\Delta t_{AB} = 46.6\pm1.1$ days, but the histogram in Fig. \ref{B1600_histo_somme} clearly shows that we cannot use the mean as the final value. The histogram has two distinct values of $\sim46$ or $\sim48$ days, so we prefer to speak of a delay of either $\Delta t_{AB} = 45.6_{-0.4}^{+1.2}$ days ($68\%$ error) or $\Delta t_{AB} = 45.6_{-0.4}^{+2.8}$ days ($95\%$ error).

\begin{figure}[htbp]
\begin{center}
\includegraphics[width=0.8\linewidth]{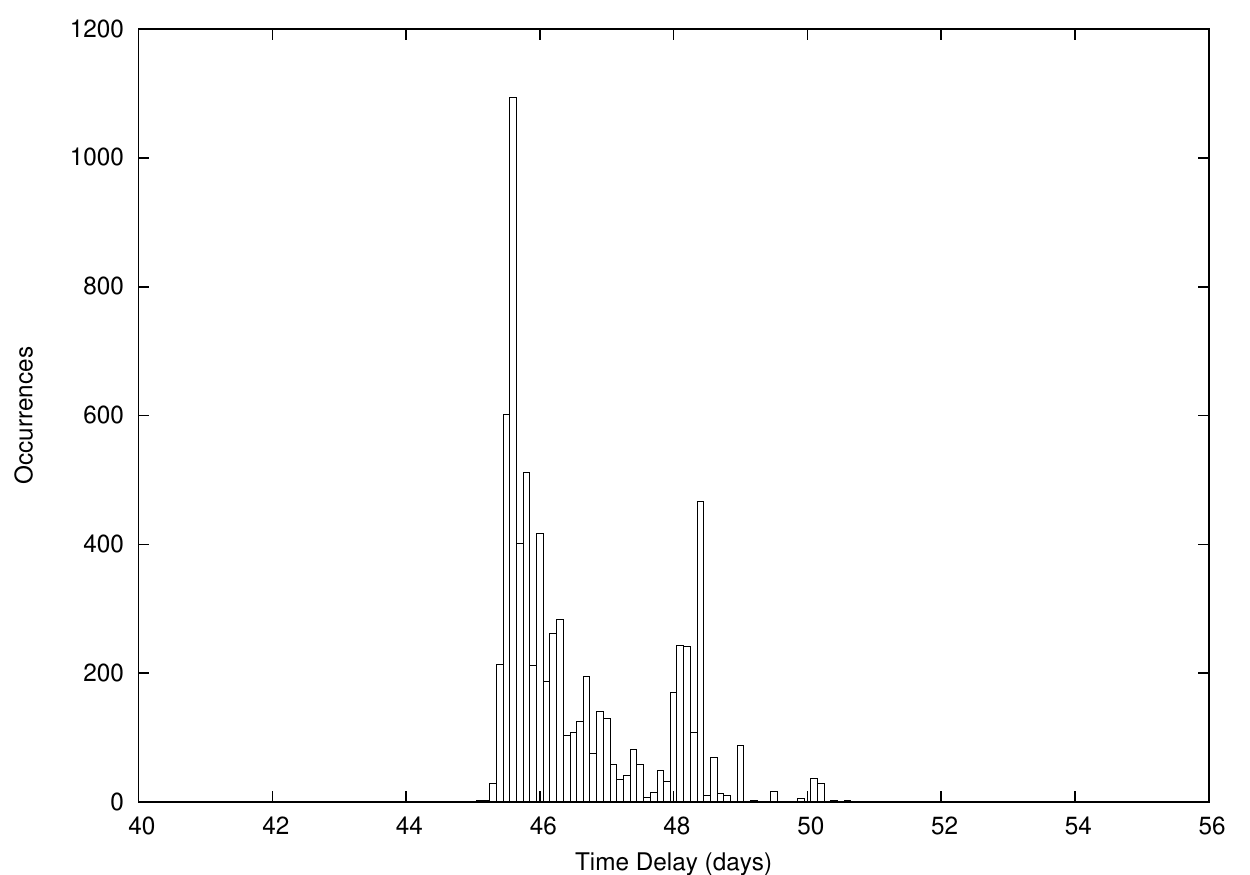}
\caption{Sum of seven histograms of 1000 runs each for B1600+434, using seven different combinations of smoothing parameters.}
\label{B1600_histo_somme}
\end{center}
\end{figure}

These results are in marginal disagreement with the final delay proposed by \citet{2000ApJ...544..117B}. However, \citet{2000ApJ...544..117B}'s final result is an average of four time delays, each of them calculated with a different method. Two of these methods also inferred a value of around $\sim48$ days. 

We could identify at least three explanations of our lower value and smaller error bars in comparison with \citet{2000ApJ...544..117B}'s time delay. The first one is the same again as for SBS 1520+530: our use of the reduced $\chi_{red}^{2}$ (see formula \ref{eqn:chi2red} in Section \ref{sec:method}) instead of the $\chi^{2}$, the latter introducing a bias towards longer delays. The second reason is the technical issue concerning the length of the model curve as explained in Section 2, which was found to be crucial for the time delay of this system. Finally, we observed that higher values of the Lagrange multiplier weighting the smoothing term seemed to lead to longer time delay values, which disappeared with lower smoothing. Taking into account these three adjustments, nearly all values around $\sim51$ days disappear from the histogram.

This is not the case for the MD method, which explains the slightly longer value of the time delay: $\Delta t_{AB} = 49.0\pm1.2$ days. Combining these results gives a delay of $\Delta t_{AB} = 47.8\pm1.2$ days. Even if these error bars imply that the time delay is very tightly constrained, we emphasize that the delay measurement is only based on $41$ data points spread over nearly two years, which gives a relatively high weight to every single data point. When adding random errors, neither of the two methods leads to a histogram with a gaussian shape. A more finely sampled light curve might remedy this situation.

 \item \textbf{CLASS B1608+656}

Light curves for this quadruply lensed system were first analysed by \citet {1999ApJ...527..498F} and subsequently improved in \citet{2002ApJ...581..823F} by adding more data. Using three observing seasons, they published time delays of $\Delta t_{BA} = 31.5_{-1}^{+2}$, $\Delta t_{BC} = 36.0\pm1.5$, and $\Delta t_{BD} = 77.0_{-1}^{+2}$ days. Their analysis is based on a simultaneous fit to data from the three seasons but treats the curves only in pairs.

\begin{figure}[htbp]
\begin{center}
\includegraphics[width=0.8\linewidth]{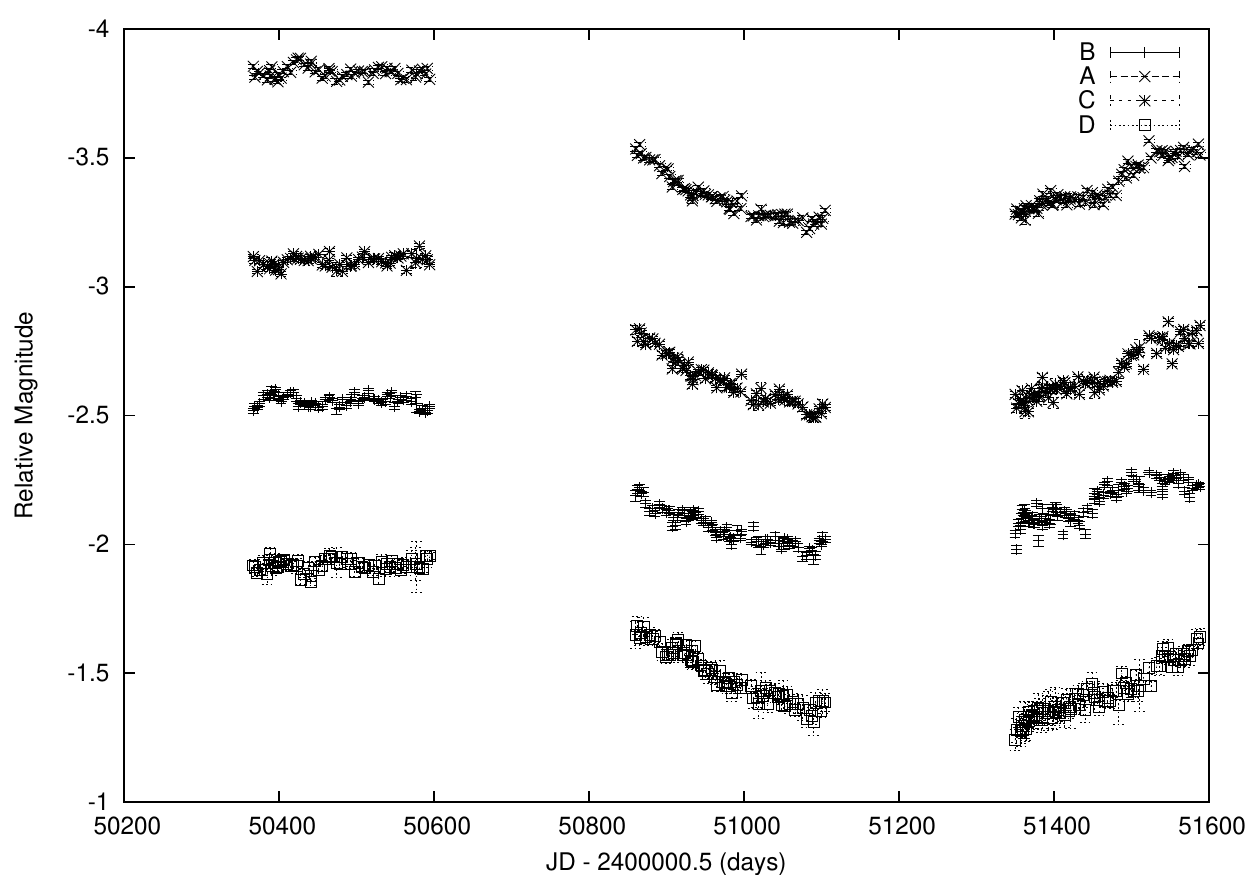}
\caption{Light curves for CLASS B1608+656. The $B$ curve has been shifted by half a magnitude for clarity.}
\label{fig:B1608}
\end{center}
\end{figure}

We performed several tests on \textbf{these light curves, which are shown in Fig.} \ref{fig:B1608}: first taking into account only the first and the third season separately (the second season not presenting useful structure), then all data for the three seasons simultaneously, using the three and four curve version of our method as described in Section \ref{sec:method}. This enables us to impose coherence between the pairs of time delays, which was not done by \citet{2002ApJ...581..823F}. The results, as illustrated in Fig. \ref{B1608_histo_3C_tout}, confirm the previously published values, within the error bars, of $\Delta t_{BA} = 30.2\pm0.9$, $\Delta t_{BC} = 36.2\pm1.1$, and $\Delta t_{BD} = 76.9\pm2.3$ days. 

\begin{figure}[htbp]
\begin{center}
\includegraphics[width=0.8\linewidth]{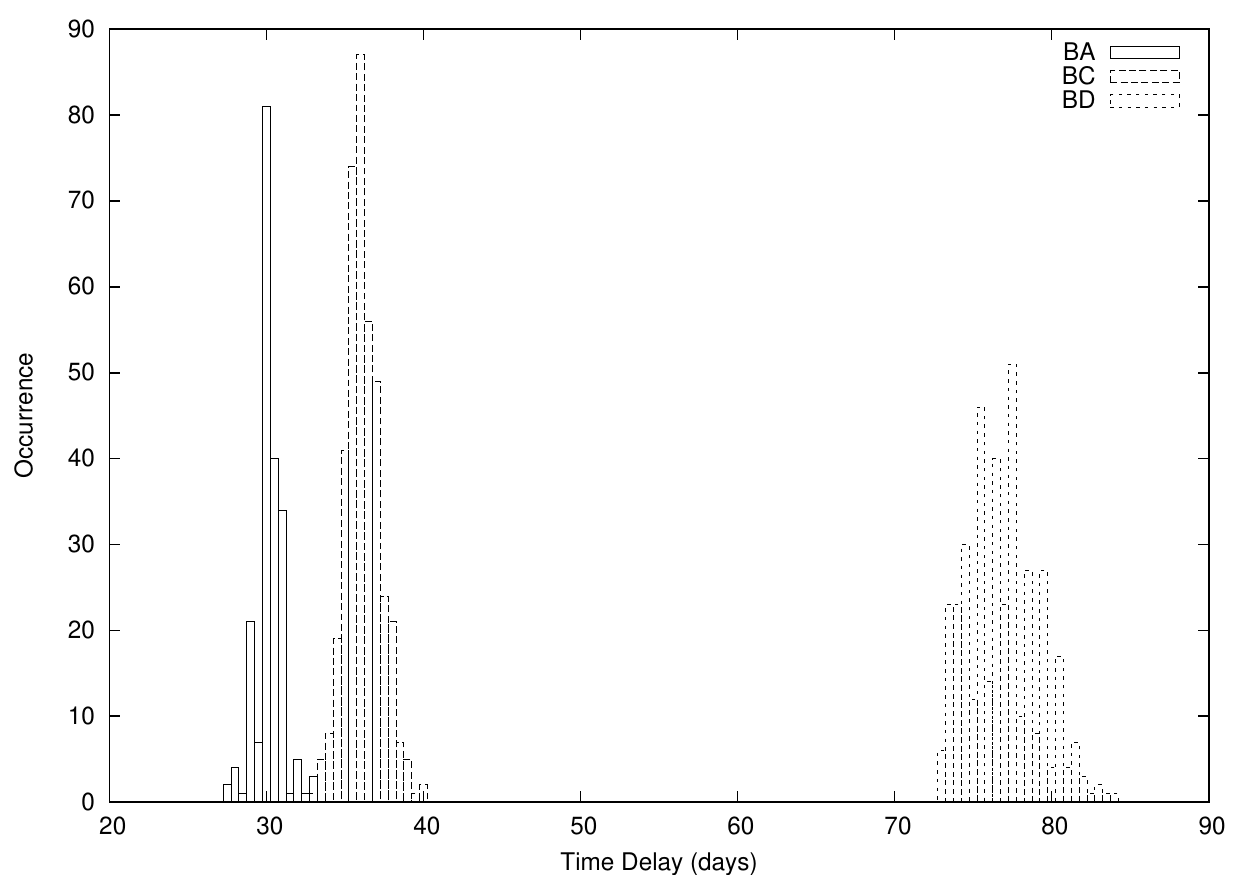}
\caption{Histograms for the three time delays in B1608+656, using two different combinations of smoothing parameters for all seasons and three out of the four curves simultaneously.}
\label{B1608_histo_3C_tout}
\end{center}
\end{figure}

One particularity deserves more attention: the value for $\Delta t_{BA}$ changes slightly \textbf{according to the seasons and the curves considered simultaneously}. 
When leaving out the second season data (featureless), $\Delta t_{BA}$ systematically converges towards $\Delta t_{BA} = 33.5\pm1.5$ days, as shown in Fig. \ref{B1608_histo_BA_4Csaison_1000}. This is consistent with \citet{2002ApJ...581..823F}, who already mentioned a time delay of $\Delta t_{AC}\sim2.5$ days. Even if this slight difference is probably due to microlensing and should be investigated in more detail, we chose to retain the final value, which is the one based on the use of all data.

\begin{figure}[htbp]
\begin{center}
\includegraphics[width=0.8\linewidth]{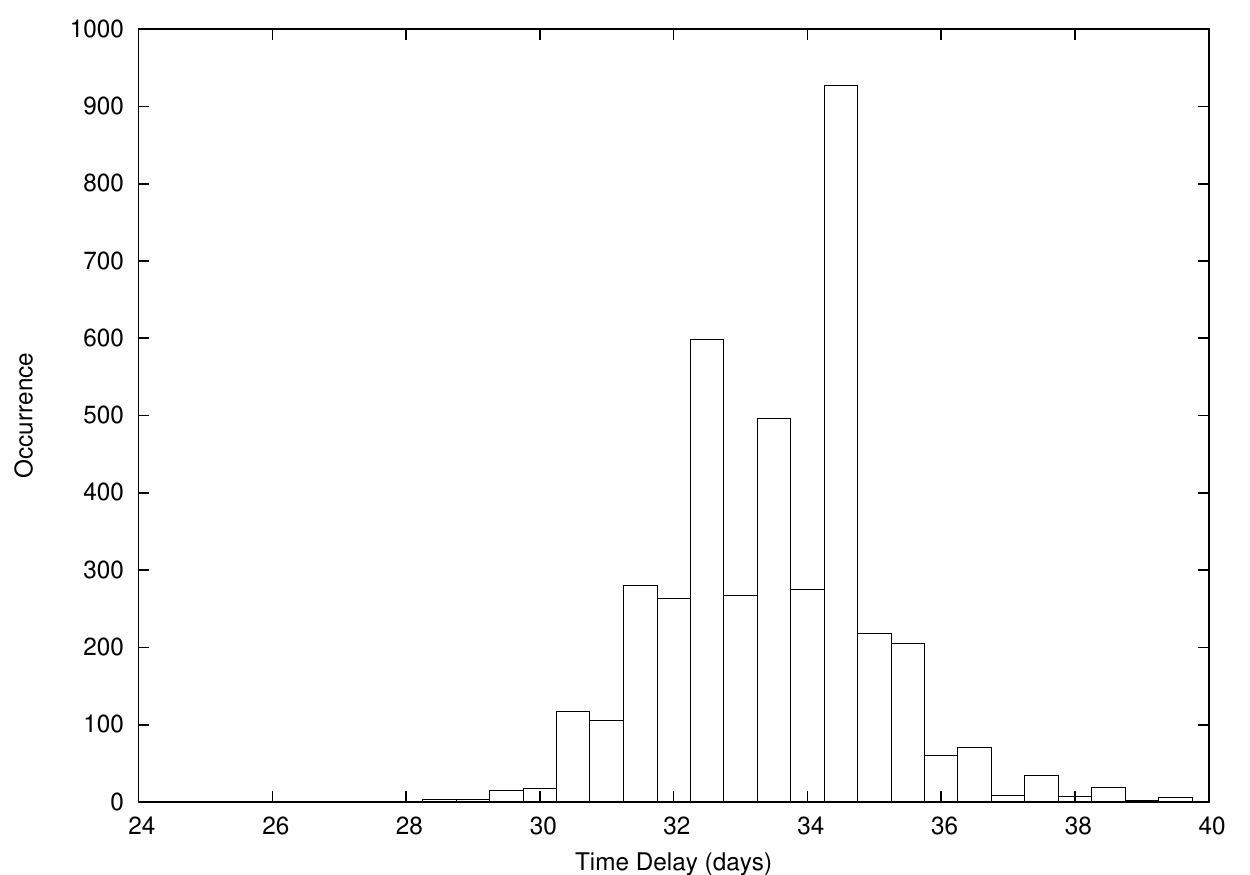}
\caption{Sum of four histograms of 1000 runs each for $\Delta t_{BA}$ in B1608+656, using two different combinations of smoothing parameters on the first and the third season.}
\label{B1608_histo_BA_4Csaison_1000}
\end{center}
\end{figure}

The MD method entirely confirms these results of $\Delta t_{BA} = 32.9\pm2.9$, $\Delta t_{BC} = 35.2\pm2.5$, and $\Delta t_{BD} = 78.0\pm3.7$ days with another indication for $\Delta t_{AC}\sim2.5$ days. Combining both methods results in the time delays of $\Delta t_{BA} = 31.6\pm1.5$, $\Delta t_{BC} = 35.7\pm1.4$, and $\Delta t_{BD} = 77.5\pm2.2$ days.

 \item \textbf{HE 2149-2745}
We reanalysed the data set made available by \citet{2002yCat..33910481B}. These data consist of two light curves, one in the $V$-band and one in the $I$-band, as shown in Fig. \ref{fig:HE2149}. \citet{2002AA...383...71B} published a time delay $\Delta t_{AB} = 103\pm12$ days where $A$ is the leading image. This delay is based on the $V$-band data, but, according to \citet{2002AA...383...71B}, agrees with the $I$-band data.

\begin{figure}[htbp]
\label{fig:V}\includegraphics[width=\columnwidth]{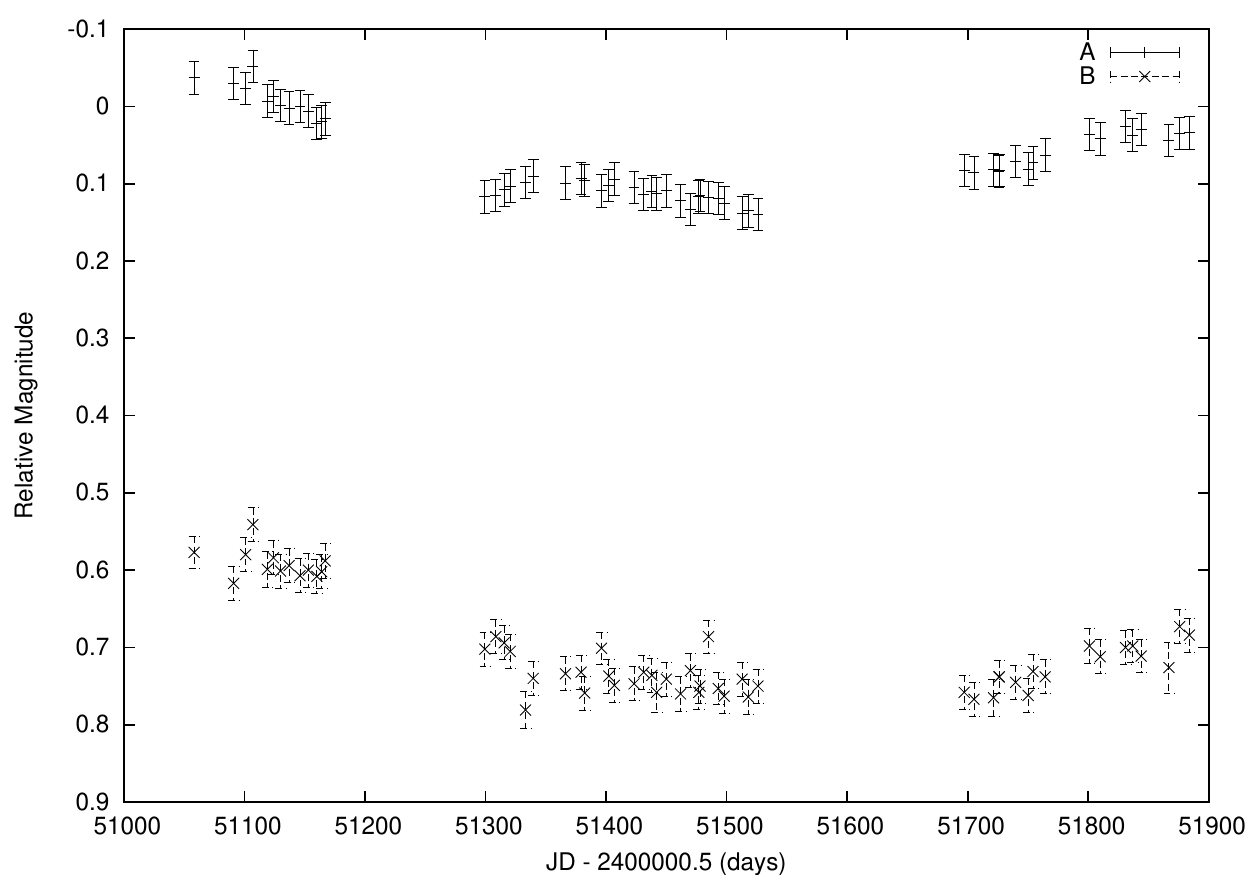}
\label{fig:I}\includegraphics[width=\columnwidth]{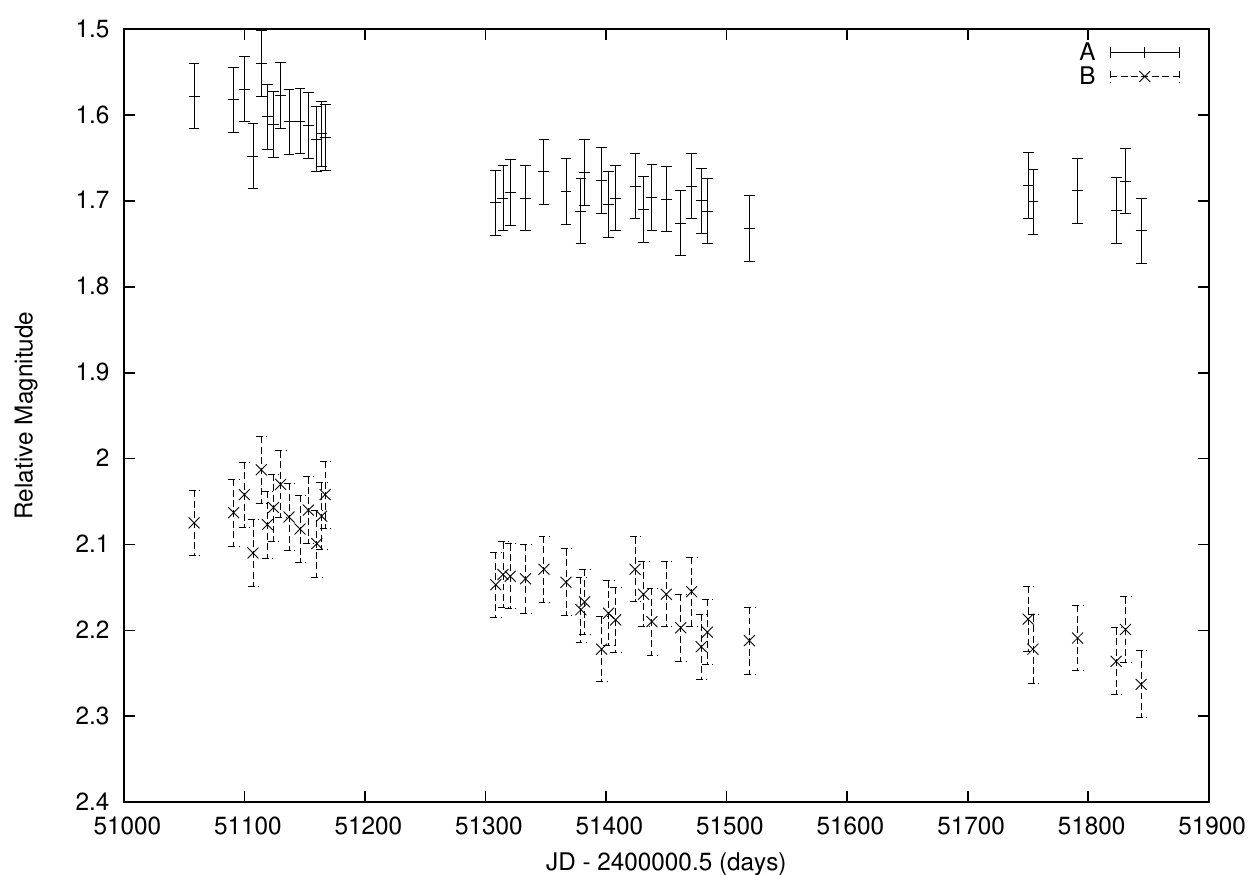}
\caption{Light curves of HE 2149-2745 in the V-band and the I-band. The $B$ curve has been shifted by one magnitude for clarity.}
\label{fig:HE2149}
\end{figure}

Our tests, based on the light curves as such, and using both methods, clearly reveal two possible delays: one around $\sim70-85$ days and one around $\sim100-110$ days. Unfortunately, the light curves of images $A$ and $B$ show little structure and hardly overlap, except for some points in the second season, when shifting them for a delay of more than $100$ days, especially in the $I$-band, which makes it very difficult to choose between the two possibilities. Moreover, once we add random errors to the model light curve and perform Monte Carlo simulations, we only obtain a forest of small peaks, spread over the entire tested range of $50-140$ days, instead of a gaussian distribution around one or two central peaks, demonstrating that these results are highly unstable. Leaving out two outlying data points in the B curve only slightly improves the situation: within the forest of peaks in Fig. \ref{HE2149_histo_somme}, those in the range $75-85$ seem to be slightly more important than those over $100$ days. Nevertheless, we cannot derive a reliable time delay from these data sets for this system. 

\begin{figure}[htbp]
\begin{center}
\includegraphics[width=0.8\linewidth]{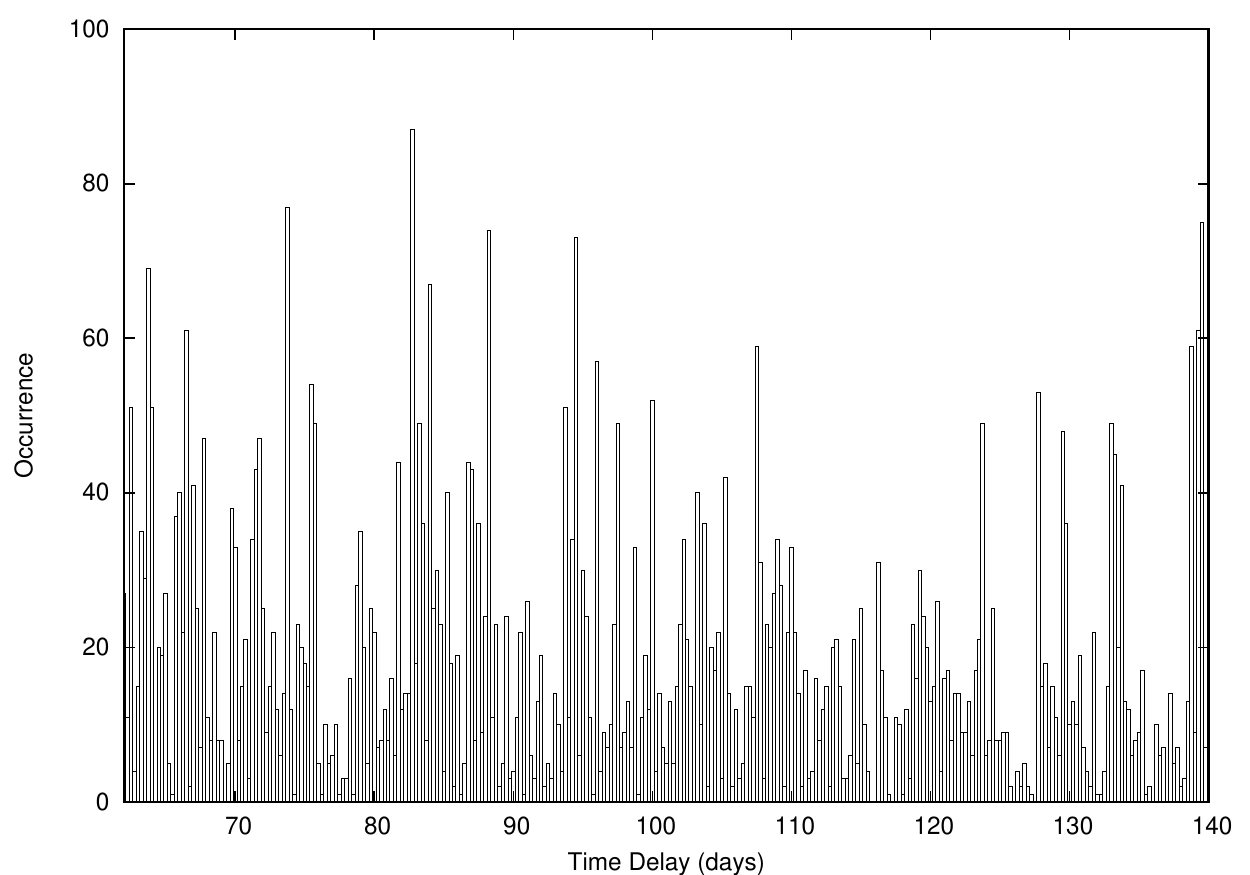}
\caption{Sum of six histograms of 1000 iterations each for HE 2149-2745 leaving out two data points, using six different combinations of smoothing parameters.}
\label{HE2149_histo_somme}
\end{center}
\end{figure}

\end{itemize}


\section{Conclusions}

We have presented an improved numerical method, the numerical model fit method, to calculate time delays in doubly or quadruply lensed quasar systems, and applied it to 11 systems for which time delays had been published previously. This allowed the validity of these time delay values to be evaluated in a coherent way. The use of a minimum dispersion method allowed us to check the independence of the results from the method. The results are summarized in Table \ref{tab:Conclusions}. 

We caution that some published time delay values should be interpreted with care: even if we have been able to confirm some values (time delays for JVAS B0218+357, HE 1104-1805, CLASS B1600+434, and for the three delays in the quadruply lensed quasar CLASS B1608+656) and give an improved value for one system, SBS 1520+530, many of the published time delays considered in our analysis have proven to be be unreliable for various reasons: the analysis is either too dependent on some data points, leads to multiple solutions, is sensitive to the addition of random errors, or is incoherent between the two different methods used.

Given the accuracy that is needed for time delays to be useful to further studies, we note that it will be necessary to perform long-term monitoring programs on dedicated telescopes to obtain high-quality light curves of lensed quasars, not only for new systems but also for the majority of the lenses in this sample, for which the time delay has been considered to be known. The COSMOGRAIL collaboration, which has been observing over 20 lensed systems for several years now, will soon be improving the time delay values for some of these systems for which the accuracy is unsatisfactory.

\begin{table*}[htbp]
\begin{center}
\begin{tabular}{lcp{3mm}ll}
System & Our Result / Comments & & Published Time Delay (days) & Reference \\
\hline
JVAS B0218+357 & $\Delta t_{AB} = 9.9_{-0.9}^{+4.0}$ & & $\Delta t_{AB} = 10.1_{-1.6}^{+1.5}$ & \citet{2000ApJ...545..578C} \\
& or & & $\Delta t_{AB} = 12\pm3$ & \citet{1996IAUS..173...37C} \\
& $\Delta t_{AB} = 11.8\pm2.3$ & & $\Delta t_{AB} = 10.5\pm0.4$ & \citet{1999MNRAS.304..349B} \\
\hline
SBS 0909+523 & unreliable & & $\Delta t_{BA} = 49\pm6$ & \citet{2008NewA...13..182G} \\
& & & $\Delta t_{BA} = 45_{-1}^{+11}$ & {\citet{2006AA...452...25U}} \\
\hline
RX J0911+0551 & 2 solutions: & & $\Delta t_{BA} = 150\pm6$ & \citet{2001PhDT.........4B}\\
& $\Delta t_{BA} \sim 146$ or $\sim 157$ & & $\Delta t_{BA} = 146\pm4$ & \citet{2002ApJ...572L..11H}\\
\hline
FBQS J0951+2635 & unreliable & & $\Delta t_{AB} = 16\pm2$ & \citet{2005AA...431..103J} \\ 
\hline
HE 1104-1805 & & & $\Delta t_{BA} = 152_{-3.0}^{+2.8}$ & \citet{2007ApJ...660..146P} \\
& & \multirow{3}{1cm}{$\left\{\rule{0mm}{6mm}\right.$} & $\Delta t_{BA} = 161\pm7$ & \citet{2003ApJ...594..101O} \\
& impossible to distinguish & & $\Delta t_{BA} = 157\pm10$ & \citet{2003AcA....53..229W} \\
& but identical within error bars & &  $\Delta t_{BA} = 162.2_{-5.9}^{+6.3}$ & \citet{2008ApJ...676...80M} \\
\hline
PG 1115+080 & dependent on method & & $\Delta t_{CA} \sim 9.4$ & \citet{1997ApJ...475L..85S} \\
&  & & $\Delta t_{CB} = 23.7\pm3.4$ & \citet{1997ApJ...475L..85S} \\
& & & $\Delta t_{CB} = 25.0_{-3.8}^{+3.3}$ & \citet{1997ApJ...489...21B}\\
\hline
JVAS B1422+231 & contradictory results & & $\Delta t_{BA} = 1.5\pm1.4$ & \citet{2001MNRAS.326.1403P} \\
& between methods: & & $\Delta t_{AC} = 7.6\pm2.5$ & \\
& BAC or CAB? & & $\Delta t_{BC} = 8.2\pm2.0$ & \\
\hline
SBS 1520+530 & $\Delta t_{AB} = 125.8\pm2.1$ & & $\Delta t_{AB} = 130\pm3$ & \citet{2002AA...391..481B} \\ 
& & & $\Delta t_{AB} = 130.5\pm2.9$ & \citet{2005AA...440...53G}\\ 
\hline
CLASS B1600+434 & $\Delta t_{AB} = 47.8\pm1.2$ & & $\Delta t_{AB} = 51\pm4$ & \citet{2000ApJ...544..117B} \\
\hline
CLASS B1608+656 & $\Delta t_{BA} = 31.6\pm1.5$ & & $\Delta t_{BA} = 31.5_{-1}^{+2}$ & \citet{2002ApJ...581..823F} \\
& $\Delta t_{BC} = 35.7\pm1.4$ & & $\Delta t_{BC} = 36.0\pm1.5$ & \\
& $\Delta t_{BD} = 77.5\pm2.2$ & & $\Delta t_{BD} = 77.0_{-1}^{+2}$ & \\
\hline
HE 2149-2745 & unreliable & & $\Delta t_{AB} = 103\pm12$ & \citet{2002AA...383...71B} \\ 
\hline
\end{tabular}
\end{center}
\caption{Summary of Time Delays for 11 Lensed Systems.}
\label{tab:Conclusions}
\end{table*}

\begin{acknowledgements}
We would like to thank Sandrine Sohy for her very precious help in programming.
This work is supported by ESA and the Belgian Federal Science Policy (BELSPO) in the framework of the PRODEX Experiment Arrangement C-90312.
\end{acknowledgements}

\bibliographystyle{aa} 
\bibliography{mybibliography}
\end{document}